\documentclass[twocolumn,showpacs,showkeys,preprintnumbers,amsmath,amssymb]{revtex4}

\usepackage{graphicx}
\usepackage{dcolumn}
\usepackage{bm}
\usepackage{amsfonts}

\begin{document}

\title{Model solution for volume reflection of relativistic
particles in a bent crystal}

\author{M.V. Bondarenco}
\email{bon@kipt.kharkov.ua}
\affiliation{%
Kharkov Institute of Physics and Technology, 1 Academic St., 61108
Kharkov, Ukraine
}%

\date{\today}

\begin{abstract}
For volume reflection process in a bent crystal, exact analytic
expressions for positively- and negatively-charged particle
trajectories are obtained within a model of parabolic continuous
potential in each interplanar interval, with the neglect of
incoherent multiple scattering. In the limit of the crystal bending
radius greatly exceeding the critical value, asymptotic formulas are
obtained for the particle mean de?ection angle in units of
Lindhard's critical angle, and for the final beam profile. Volume
re?ection of negatively charged particles is shown to contain
effects of rainbow scattering and orbiting, whereas with positively
charged particles none of these effects arise within the given
model. The model predictions are compared with experimental results
and numerical simulations. Estimates of the volume re?ection mean
angle and the final beam profile robustness under multiple
scattering are performed.
\end{abstract}

\pacs{61.85.+p, 29.27.-a, 45.10.-b}
\keywords{volume reflection; harmonic inter-planar potential; rainbow scattering; orbiting}
\maketitle


\section{Introduction}\label{sec:intro}
The volume re?ection is an effect of deflection of high- energy
charged particles upon their \emph{over-barrier} (nonchanneled)
passage through a planarly oriented bent crystal. The effect arises
when the crystal bending radius $R$ greatly exceeds the critical
value $R_c$. That condition is the same as the Tsyganov's one for
the possibility of channeling in a bent crystal \cite{Tsyganov}, but
the particle motion regime in the crystal yet depends on the
particle entry angle relative to the active atomic planes.When this
angle is much larger than the critical value $\theta_c$, then,
moving in the continuous potential of bent planes, conserving the
particle full transverse (radial) energy, the particles are rarely
captured into channels (via incoherent scattering on atomic
electrons and nuclei), and are mostly deflected elastically through
the volume reflection mechanism. Curiously, the latter deflection
proceeds to the side \emph{opposite} to that of the crystal bending;
the value of the deflection angle is of the order of critical angle
$\theta_c$. Furthermore, the particle beam after deflection remains
fairly well collimated, i.e., its angular dispersion keeps much
smaller than the mean deflection angle. That phenomenon was
discovered in numerical simulations two decades ago  \cite{Tar-Vor}
and recently verified experimentally \cite{Scandale,Scandale-neg}.
Nowadays it is considered to be an option for beam collimation and
partial extraction at ultrarelativistic charged-particle
accelerators \cite{Tar-Scand,Bir-Breese,mult-VR}.

To a good accuracy, the particle dynamics in the volume reflection
problem is classical \cite{Lindhard} and reduces to classical
particle motion in a cylindrically symmetrical continuous potential
of bent atomic planes. Therewith, granted the angular momentum
conservation relative to the active crystallographic plane bending
axis, the final deflection angle is expressible in the standard way
as an integral over the radial coordinate from an inverse square
root function involving the potential [see
Eq.~(\ref{thetarefl-simpler}) below]. That approximation served as a
starting point for a number of numerical studies
\cite{Tar-Vor,Maisheev}.

Although the described computational problem seems to be
sufficiently simple, it is aggravated by the presence of several
parameters: the ratio $R/R_c$, initial and final particle variables
(impact parameter, the angles of incidence and deflection). The
dependencies on all those parameters involve singularities, which in
general are better dealt with by analytic techniques than by
numerical ones. Besides that, as long as for practice mostly
interesting is the case $R\gg R_c$, it would be instructive to
evaluate the asymptotic behavior of all relevant observables in the
formal limit $R/R_c\to\infty$, including next-to-leading order
corrections in the small parameter $R_c/R$. But since volume
reflection depends on the particle dynamics not in one but in
several inter-planar intervals, for feasibility of its global
analytic description one rather needs some model.

A valuable opportunity for realistic model building is that the
inter-planar potential in a silicon crystal, at least in the
orientation (110), is fairly close to parabolic shape over the
\emph{entire} inter-planar interval (see, e. g., \cite{Bagli}). A
parabolic (harmonic) potential, i.e., a linear oscillator, permits a
simple solution for the particle trajectory within a single
interplanar interval. The next problem is to connect solutions on
the boundaries of the adjacent intervals. It may appear nontrivial,
but it is feasible to do that transitively, i.e., simultaneously for
an arbitrary number of the adjacent intervals. Thereby we obtain a
completely solvable model capturing basic features of the volume
re?ection, except the effects of incoherent multiple
scattering.Moreover, we are able to derive not only the de?ection
angle, but also an expression for the whole trajectory, which
further on may be used for description of inelastic processes, such
as volume capture or electromagnetic radiation.

In the present work we will deliver a solution for the posed model
problem. The plan of the article is as follows. In
Sec.~\ref{sec:1stinterv} we describe the procedure of solution
connection between adjacent inter-planar intervals, demonstrating
that the problem reduces to elementary trigonometry. The particle
trajectory is expressed as an explicit function of inter-planar
interval order number (not involving a recursive procedure), for
arbitrary ratio $R/R_c$. In Sec.~\ref{sec:volrefl}, from the
obtained solution for trajectory, we derive the particle final
deflection angle, which comes as a sum of inverse trigonometric (for
positively charged particles) or hyperbolic (for negatively charged
particles) functions. In Sec.~\ref{sec:largeR} we scrutinize the
limit $R\gg R_c$, most interesting in relation to volume reflection
and practical applications, first for positively, then for
negatively charged particles. In the generic expression for the ?nal
deflection angle, we find a possibility to replace the sums involved
by integrals (via the Euler-Maclaurin formula), and do the latter
ones in a closed form. As a result, we arrive at sufficiently simple
asymptotic formulas for the de?ection angle dependence on all the
variables. The impact parameters are thereupon analytically averaged
over, and the experimentally observable scattering differential
cross section is obtained for positive and negative particles. In
Sec.~\ref{sec:pert-limit} we examine the opposite limit $R\ll R_c$.
In Sec.~\ref{sec:optimiz} we provide estimates of optimal crystal
and initial beam parameters for beam complete deflection or for
experimental investigation of the final beam profile features. A
summary is given in Sec.~\ref{sec:summary}.

\section{Particle trajectory in a bent crystal}\label{sec:1stinterv}

\subsection{Initial conditions}

\begin{figure}
\includegraphics{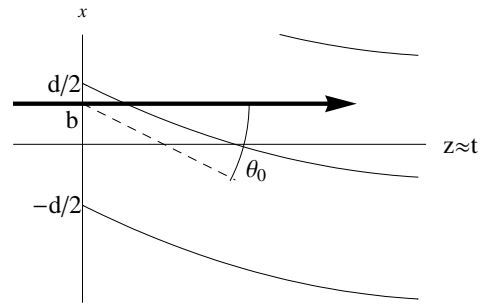}
\caption{\label{fig:passage} Coordinates describing the bent crystal
geometry (circle segments are the bent atomic planes) and the
particle entrance to the crystal (thick arrow). Not to scale. For
details see text.}
\end{figure}

The usual geometry of experiments on volume reflection implies
sending a charged particle beam normally to a thin \footnote{The
crystal has to be thin in order to avoid the strong influence of
multiple scattering, but still it may be thick enough for the volume
reflection to occur within the crystal volume and be independent of
the boundaries. We shall quantify this requirement later on.},
weakly bent crystal plate. The practically unavoidable slight
curvature of the crystal boundary thereat is of minor consequence,
since the main contribution to the particle reflection angle comes
from a vicinity of some point in the depth of the crystal. For
definiteness and to establish an easy connection with the particle
impact parameter in the initial (perfectly parallel \footnote{We can
turn to the issue of the initial beam divergence after we derive the
scattering differential cross-section.}) beam, let us consider a
particle incident along the $z$ axis on a crystal whose front face
is a \emph{perfect plane}, located at $z=0$. As for the crystal rear
face, for our purposes in this paper we may leave it unspecified at
all, as if the crystal was infinitely thick, but transparent. Then,
let $\theta_0$ ($0<\theta_0\ll1$) be the angle of inclination of
crystalline planes to the $z$-axis at the crystal front face (see
Fig.~\ref{fig:passage}), and let $x$-axis, perpendicular to $Oz$,
point in the direction of the crystal bend. Moving at small angles
to the crystal planes, the particle interacts most strongly
(coherently) with the averaged, so-called continuous inter-planar
potential \cite{Lindhard}, which induces a force with dominant
$x$-component (yet slowly dependent on $z$) \footnote{At the very
entrance to the crystal the force $z$-component may become
comparable to $x$-component, in order to preserve the force
nonvorticity but edge effects may certainly be neglected for a
deeply penetrating particle.}. Along the $y$ coordinate there is a
translational invariance, ensuring the particle momentum
$y$-component conservation.

In monocrystals of not too heavy chemical elements, in particular
for silicon (lattice of diamond-type), oriented by its (110) plane
close to the direction of the beam,
the continuous potential in each inter-planar interval may closely
be approximated by a quadratic function, with an accuracy
$\lesssim20\%$ \footnote{The condition thereof is that the
Thomas-Fermi radius of the material atom be commensurable with the
inter-atomic distance half width. In silicon that is the case we
have. As for crystals of heavier elements such as tungsten, thereat
the atomic radius is appreciably smaller, and the parabolic
approximation for the inter-planar continuous potential is poor.}.
That entails a linear equation of motion for the classical
 \footnote{Standardly \cite{Lindhard}, for a
high-energy particle interacting with an oriented crystal, the
particle wavelength shortness on the atomic scale makes the particle
dynamics essentially classical, but yet non-perturbative, given the
small angle of particle motion relative to a crystallographic
direction and thus coherent action of atomic forces over long
distances. Still, quantum effects may be viable in a special case
when the particle transverse energy is very close to the height of
an atomic potential barrier (the author is indebted to A.V. Shchagin
for pointing this out in private conversation), but in any case, the
classical calculation has to pave the way.} ultra-relativistic
\footnote{Equation (\ref{diff}) in itself may apply to a
non-ultra-relativistic motion, too, provided $E$ includes the
particle rest energy (so, in the non-relativistic limit $E\to m$).
But physically, in order not to complicate the analysis, we confine
ourselves to (the most important) ultra-relativistic case in the
present paper.} particle:
\begin{equation}\label{diff}
    \ddot{x}=\frac{2F_{\max}}{Ed}\left(-x+x_0\right),
\end{equation}
\begin{equation}\label{}
    t\approx z\qquad (c=1,\,\, \mathrm{small\, angle\, motion}),
\end{equation}
where $x_0$ is the midpoint of the inter-planar interval,  $d$ the
inter-planar distance, $E$ the particle energy, and $F_{\max}$ the
force acting on the particle at the edge of the inter-planar
interval $x-x_0=-\frac d2$. For positively charged particles
$F_{\max}$ is positive, whereas for negatively charged particles it
is negative. Note that the force and the particle energy enter
equation (\ref{diff}) only through the ratio
\begin{equation}\label{Rc-def}
    \frac E{|F_{\max}|}=R_c,
\end{equation}
known as the Tsyganov critical radius \cite{Tsyganov}. The natural
time unit in channeling-related phenomena is
\begin{equation}\label{tau}
    \tau=\sqrt{\frac{Ed}{2|F_{\max}|}}\equiv\sqrt{\frac{R_cd}2}
\end{equation}
($2\pi\tau$ has the meaning of positively charged particle
channeling period, although herein we deal not with channeling but
with an over-barrier motion).

One and the only consequence of the crystal bending is that $x_0$ in
Eq.~(\ref{diff}) acquires dependence on the longitudinal coordinate
$z$, which for ultra-relativistic motion under small angles to $Oz$
may be equated to the current time $t$:
\begin{equation}\label{}
    x_0=x_0(z\approx t)\qquad (\mathrm{the\, crystal\, bend\, function}).
\end{equation}
In application to volume reflection, we are interested in the
uniform bending of the crystal, at which $x_0(t)$ describes a
circular arc of a small opening angle. That small arc may equally
well be approximated by a parabola, and hence $x_0(t)$ is determined
by the equation
\begin{equation}\label{Rc-def}
    x_0(t)=-\theta_0t+\frac{t^2}{2R}\quad (\mathrm{uniformly\, bent\, crystal}),
\end{equation}
where $R$ is the atomic plane bending radius (without the loss of
generality one may let $x_0(0)=0$ -- see Fig.~\ref{fig:passage}).
Inserting (\ref{Rc-def}) to (\ref{diff}), and implementing
(\ref{tau}), we get the particle equation of motion in the first
inter-planar interval:
\begin{equation}\label{}
    \ddot{x}=\pm\frac1{\tau^2}\!\left(\!-x-\theta_0t+\frac{t^2}{2R}\right)\quad\left\{\begin{array}{c}
                                                                                                     \mathrm{pos.\,charged\,particles} \\
                                                                                                     \mathrm{neg.\,charged\,particles}
                                                                                                   \end{array}
    \right\}.
\end{equation}
Initial conditions for $x(t)$ stand as
\begin{eqnarray}
    x(0)&=&b,\label{init-x}\\
    \dot{x}(0)&=&0,\label{init-xdot}
\end{eqnarray}
where $b$, restricted to the interval
\begin{equation}\label{b-interval}
    -\frac d2\leq b\leq\frac d2,
\end{equation}
is the impact parameter measured from the middle of the interval

The equations of motion further simplify in terms of the
``subtracted radius" variable
\begin{equation}\label{}
    r(t)=-x(t)-\theta_0t+\frac{t^2}{2R},
\end{equation}
becoming
\begin{equation}\label{r-eq-1st}
    \ddot{r}=\frac{\delta\mp r}{\tau^2}\qquad \left(\mathrm{in}\,\, -\frac d2\leq r\leq\frac
    d2\right),
\end{equation}
where
\begin{equation}\label{delta-def}
    \delta=\frac{\tau^2}R.
\end{equation}
Thus, $\pm\delta$ is the spatial shift of the oscillator equilibrium
position due to the crystal bend, i. e., due to the centrifugal
force, which in the present small-angle approximation, presuming
condition $r\ll R$ within the weakly bent crystal, is treated as
virtually independent of the subtracted radius $r$ (cf.
\cite{Tar-Vor}). For $r(t)$, the initial conditions
(\ref{init-x}-\ref{init-xdot}) translate to
\begin{eqnarray}
    r(0)&=&-b,\label{init-r}\\
    \dot{r}(0)&=&-\theta_0.\label{init-rdot}
\end{eqnarray}

Generic solution of Eq.~(\ref{r-eq-1st}) reads:
\begin{equation}\label{r0}
    r_0(t)=\pm\delta-A_0\left\{\begin{array}{c}
                            \sin \\
                            \sinh
                          \end{array}
    \right\}\left(\frac{t}\tau+\varphi_0\right).
\end{equation}
(Here and henceforth upper signs and figures refer to positively
charged particles, and lower ones -- to negatively charged
particles.) Matching initial conditions
(\ref{init-r}-\ref{init-rdot}) allows one to determine constants
$A_0$ and $\varphi_0$:
\begin{equation}\label{}
    A_0=\sqrt{\tau^2\theta_0^2\pm(b\pm\delta)^2},
\end{equation}
\begin{equation}\label{}
    \varphi_0=\left\{\begin{array}{c}
                            \arcsin \\
                            \mathrm{arsinh}
                          \end{array}
    \right\}\frac{b\pm\delta}{A_0}
\end{equation}
(the sign of $A_0$ must be chosen  positive in order that $\dot
r(0)$ at $\theta_0>0$, according to (\ref{init-rdot}), was
negative).

Further on, solution (\ref{r0}) is to be connected with solutions in
the subsequent inter-planar intervals. Importantly, since these
solutions have to be connected at a definite $r$, the condition of
connection will not depend on the current phase of the harmonic
motion, such as $\varphi_0$ in Eq.~(\ref{r0}), as we are going to
show.

\subsection{Connection of solutions through
interval borders}\label{subsec:1stinterv}

\begin{figure}
\includegraphics[width=85mm]{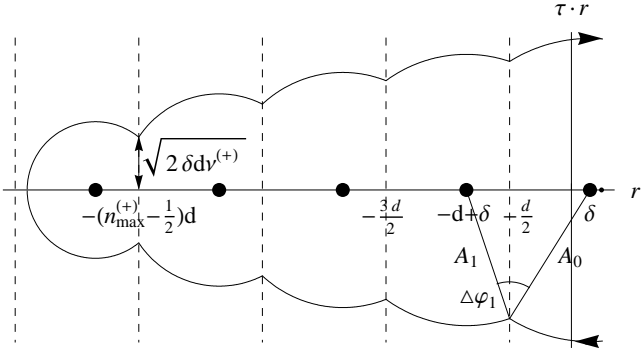}
\caption{\label{fig:circles} Solid curve -- the phase-space (the
subtracted radius $r$ vs. the radial velocity $\dot r$) trajectory
for positively charged particles, under condition $\delta<\frac d2$.
Dashed vertical lines signify the positions of bent atomic planes
(definite $r$). Thick dots indicate centers of the trajectory
circular segments. Vertex angle $\triangle\varphi_1$ (and similarly
all other $\triangle\varphi_n$) may be interpreted as a geometric
sum of vertex angles in a pair of right triangles having a common
cathetus, and with the second catheti equal $\frac d2+\delta$,
$\frac d2-\delta$, and the hypotenuses $A_0$, $A_1$.}
\end{figure}

Moving along trajectory (\ref{r0}), the particle will cross the next
inter-planar interval border $r=-\frac d2$ at an instant
\begin{equation}\label{}
    \frac{t_1}\tau=\left\{\begin{array}{c}
                            \arcsin \\
                            \mathrm{arsinh}
                          \end{array}
    \right\}\frac{\frac d2\pm\delta}{A_0}-\varphi_0
\end{equation}
(inferred from (\ref{r0}) by letting $r=-\frac d2$ and solving for
$t$). At this instant, the equation of particle motion turns to
\begin{equation}\label{eee}
    \ddot{r}=\frac{\delta\mp (d+r)}{\tau^2}\qquad \left(\mathrm{in}\,\,-\frac{3d}2\leq r\leq-\frac
    d2\right).
\end{equation}
That is the same harmonic oscillator, only with an altered
equilibrium position, and general solution of (\ref{eee}) may be
written as
\begin{equation}\label{r1}
    r_1(t)=\pm\delta-d-A_1\left\{\begin{array}{c}
                            \sin \\
                            \sinh
                          \end{array}
    \right\}\left(\frac
    t\tau+\varphi_0+\triangle\varphi_1\right).
\end{equation}
Values of new constants $A_1$, $\bigtriangleup\varphi_1$ are now to
be determined from the continuity of $r(t)$ and $\dot{r}(t)$ at the
border point $r=-\frac d2$. That can be done without formally
solving the system of two equations. First, compare two integrals of
motion
\begin{eqnarray}
  A_1^2 &=& \tau^2\dot{r}_1^2\pm\left(\pm\delta-d-r_1\right)^2 \label{A12}\\
  A_0^2 &=& \tau^2\dot{r}^2_0\pm\left(\pm\delta-r_0\right)^2 \label{A02}
\end{eqnarray}
(related to transverse energy)  in their common point, where
$r_0=r_1=-\frac d2$, $\dot{r}_0=\dot{r_1}$. Subtracting (\ref{A02})
from (\ref{A12}), one gets $A_1^2=A_0^2-2\delta d$, i. e.
\begin{equation}\label{}
    A_1=\sqrt{A_0^2-2\delta d}.
\end{equation}
Thereupon, the phase shift $\bigtriangleup\varphi_1$ is sought from
a condition $r_1(t_1)=-\frac d2$. One finds:
\begin{equation}\label{}
    \triangle\varphi_1=-\left\{\begin{array}{c}
                            \arcsin \\
                            \mathrm{arsinh}
                          \end{array}
    \right\}\frac{\frac d2\pm\delta}{A_0}-\left\{\begin{array}{c}
                            \arcsin \\
                            \mathrm{arsinh}
                          \end{array}
    \right\}\frac{\frac d2\mp\delta}{A_1}
\end{equation}
(for a geometric interpretation of this relation for positive
particles -- see Fig.~\ref{fig:circles}). As we had expected,
neither $A_1$, nor $\triangle\varphi_1$ depends on $\varphi_0$.

At each next border connection of the solutions is implemented in
exactly the same way. Writing in the $n$-th interval
\begin{eqnarray}\label{rn}
    r_n(t)=\pm\delta-nd-\!A_n\!\left\{\begin{array}{c}
                            \sin \\
                            \sinh
                          \end{array}
    \right\}\!\left(\frac
    t\tau+\varphi_0+\!\sum_{m=1}^n\!\triangle\varphi_m\right)\!,\nonumber\\
\end{eqnarray}
\[
\left(-\frac d2-nd\leq r\leq\frac d2-nd,\quad t_n\leq t\leq
t_{n+1}\right)
\]
the generic amplitude is found as
\begin{eqnarray}\label{An}
    A_n&=&\sqrt{A^2_{n-1}-2\delta d}=\sqrt{A_0^2-2n\delta d}\nonumber\\
    &=&\sqrt{\tau^2\theta_0^2\pm(b\pm\delta)^2-2n\delta
    d}.
\end{eqnarray}
and the phase shift is deduced to be
\begin{equation}\label{deltaphin}
    \triangle\varphi_n=-\left\{\begin{array}{c}
                            \arcsin \\
                            \mathrm{arsinh}
                          \end{array}
    \right\}\frac{\frac d2\pm\delta}{A_{n-1}}-\left\{\begin{array}{c}
                            \arcsin \\
                            \mathrm{arsinh}
                          \end{array}
    \right\}\frac{\frac
    d2\mp\delta}{A_n},
\end{equation}
where the amplitudes in the denominators must be treated as already
known, by (\ref{An}). The instants of border passage can also be
evaluated:
\begin{eqnarray}\label{tn}
    \frac{t_n}\tau&=&\sum_{m=1}^n\left\{\begin{array}{c}
                            \arcsin \\
                            \mathrm{arsinh}
                          \end{array}
    \right\}\frac{\frac
    d2\pm\delta}{A_{m-1}}+\sum_{m=1}^{n-1}\left\{\begin{array}{c}
                            \arcsin \\
                            \mathrm{arsinh}
                          \end{array}
    \right\}\frac{\frac
    d2\mp\delta}{A_m}\nonumber\\
    &\,&-\left\{\begin{array}{c}
                            \arcsin \\
                            \mathrm{arsinh}
                          \end{array}
    \right\}\frac{b\pm\delta}{A_0}.\qquad\qquad(r_{n-1}\to r_n)
\end{eqnarray}

\begin{figure}
\includegraphics[width=85mm]{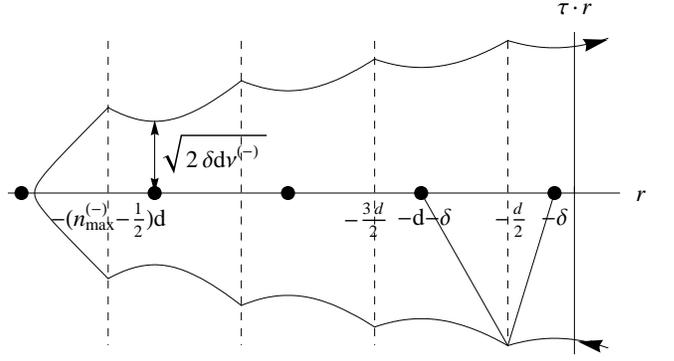}
\caption{\label{fig:hyperb} The same as Fig.~\ref{fig:circles}, but
for negatively charged particles, assuming condition $\delta<\frac
d2$ (see Sec.~\ref{sec:neg-part}).}
\end{figure}

One caution is that amplitudes $A_n$ should not be regarded as a
measure of the particle spatial wiggling in each interval. As
Figs.~\ref{fig:circles}, \ref{fig:hyperb} indicate, the trajectory
swinging \emph{enhances} as the particle penetrates deeper into the
crystal, whereas amplitudes $A_n$, to the contrary, \emph{decrease}.
There is no contradiction hereat because for most of the intervals
traversed the intra-channel oscillation period $2\pi\tau$ is much
greater than the time of particle passage across the interval, thus
the particle is far from making a full oscillation in each interval,
anyway. In fact, the lower is the amplitude $A_n$ compared to the
interval length (along the particle motion direction), the stronger
warp of the trajectory on this interval may occur (see
Figs.~\ref{fig:circles}, \ref{fig:hyperb}).

\section{Particle reflection}\label{sec:volrefl}
\subsection{Reflection conditions for positive particles}
It is clear that the decrease of amplitudes (\ref{An}) can not
continue indefinitely, because eventually arguments of the arcsines
in (\ref{deltaphin}) shall exceed unity (that happens sooner than
the radicand in (\ref{An}) becomes negative). This merely signals
that the particle can not reach the next inter-planar interval. The
particle will continue the harmonic motion until it hits the
previous interval, then proceeds moving outwards in the radial
variable in the same but reverse way, and on the exit from the
crystal it will emerge as a deflected beam.

Let us evaluate the order number $n_{\max}^{(+)}$ of the reflection
interval. If for some $n$ inequality $\frac d2+\delta\leq A_{n-1}$
is met, then it also entails $\frac d2-\delta\leq A_n$, so arguments
of all the arcsines in (\ref{tn}) are less than unity. So,
$n_{\max}^{(+)}$ is the largest integer yet allowing for $\frac
d2-\delta\leq A_{n_{\max}^{(+)}}$. Through (\ref{An}), that
condition determines the reflection interval order number:
\begin{equation}\label{nmax+def}
    n_{\max}^{(+)}=\left\lfloor\frac{\tau^2\theta_0^2+(b+\delta)^2-\left(\frac d2-\delta\right)^2}{2\delta
    d}\right\rfloor,
\end{equation}
where the lower-corner brackets $\lfloor\ldots\rfloor$ designate the
integer part of a number ($\lfloor a\rfloor\leq a$). If
$\tau\theta_0\gg d$, variation of $n_{\max}(b)$ is much smaller than
its mean value
\begin{equation}\label{nmax+mean}
    n_{\max}\sim\frac{\tau^2\theta_0^2}{2\delta
    d}=\frac{R}{2d}\theta_0^2
\end{equation}
(essentially valid for negative particles as well -- see
Eq.~(\ref{nmax-unif}) below). It is instructive to notice that
\[
\frac d2-\delta\leq A_{n_{\max}^{(+)}}< \frac d2+\delta.
\]

Toward the volume reflection problem, we are interested in finding
the total reflection angle $\theta_{\mathrm{refl}}$, half of which,
by symmetry reasons, amounts to deflection angle of the trajectory
in the reflection point $t=t_{\mathrm{refl}}$ in which
\begin{equation}\label{}
    \dot
r(t_{\mathrm{refl}})=0,
\end{equation}
i. e.,
\begin{equation}\label{thetarefl}
    \frac12\theta_\mathrm{refl}\simeq\dot{x}(t_\mathrm{refl})=-\theta_0+\delta\frac{t_\mathrm{refl}}{\tau^2}
\end{equation}
(more exactly -- see Eq.~(\ref{thetavolrefldef}) below)
\footnote{Strictly speaking, the trajectory will not be exactly
symmetric with respect to $t_{\mathrm{refl}}$ because the distances
from $t_{\mathrm{refl}}$ to the crystal boundaries are in general
different. However, contributions to $\theta_\mathrm{refl}$ from
crystal regions away from $t_{\mathrm{refl}}$ are supposed to
decrease sufficiently rapidly, and one expects existence of a
``thick crystal limit" of $\theta_{\mathrm{refl}}$, relevant in
actual practice -- see Sec.~\ref{sec:largeR}. In this paper, we
content ourselves only to the ``volume" contribution
(\ref{thetarefl}) to $\theta_{\mathrm{refl}}$ and do not study any
boundary effects. As we shall see later (Sec.~\ref{sec:largeR} and
Appendix), however, the omission of boundary effects requires
certain care.}. To evaluate the right-hand side of
(\ref{thetarefl}), one only needs to know the value of
$t_\mathrm{refl}$. The latter is found from solving equation $\dot
r(t_{\mathrm{refl}})=0$:
\begin{eqnarray}\label{trefl}
    \frac{t_\mathrm{refl}}\tau&=&\frac{t_{n_{\max}^{(+)}}}\tau+\frac\pi2\nonumber\\
    &=&\frac\pi2+\sum_{n=1}^{n_{\max}^{(+)}}\left(\arcsin\frac{\frac d2+\delta}{A_{n-1}}+\arcsin\frac{\frac d2-\delta}{A_n}\right)\nonumber\\
    &\,&-\arcsin\frac{b+\delta}{A_0}.
\end{eqnarray}
The largest contribution to the emerging sum comes from the terms
$n\sim n_{\max}^{(+)}$ (where denominators $A_n$ are smallest), so
it may be more convenient to revert here the summation order.
Introducing a useful parameter
\begin{equation}\label{nu+def}
    \nu^{(+)}=\textbf{\Bigg\{}\frac{\tau^2\theta_0^2+(b+\delta)^2-\left(\frac d2-\delta\right)^2}{2\delta
    d}\mathbb{\Bigg\}}_\mathrm{f},
\end{equation}
with braces $\{\ldots\}_\mathrm{f}$ to indicate the fractional part
($0\leq\nu^{(+)}<1$), one recasts (\ref{trefl}) as
\begin{subequations}\label{trefl2-c}
\begin{eqnarray}\label{trefl2}
    \frac{t_\mathrm{refl}}\tau&=&\frac\pi2-\arcsin\frac{b+\delta}{\sqrt{\left(\frac d2-\delta\right)^2+2\left(\nu^{(+)}+n_{\max}^{(+)}\right)\delta d}}\nonumber\\
    &\,&+\sum_{n=0}^{n_{\max}^{(+)}-1}\Bigg(\!\arcsin\frac{\frac d2+\delta}{\sqrt{\left(\frac d2+\delta\right)^2+2\left(\nu^{(+)}+n\right)\delta d}}\nonumber\\
    &\,&+\arcsin\frac{\frac d2-\delta}{\sqrt{\left(\frac d2-\delta\right)^2+2\left(\nu^{(+)}+n\right)\delta
    d}}\Bigg).
\end{eqnarray}
Equivalently, using the identity
$\arcsin\frac1{\sqrt{1+\eta}}=\mathrm{arccot}\sqrt{\eta}$, one can
write
\begin{eqnarray}\label{trefl3}
    \frac{t_\mathrm{refl}}\tau=\frac\pi2-\arcsin\frac{b+\delta}{\sqrt{\left(\frac d2-\delta\right)^2+2\left(\nu^{(+)}+n_{\max}^{(+)}\right)\delta d}}\quad\nonumber\\
    +\!\!\sum_{n=0}^{n_{\max}^{(+)}\!-\!1}\!\!\!\Bigg(\!\mathrm{arccot}\frac{\!\sqrt{2(\nu^{(+)}\!\!+\!n)\delta d}}{\frac
    d2+\delta}\!
    +\!\mathrm{arccot}\frac{\!\sqrt{2(\nu^{(+)}\!\!+\!n)\delta d}}{\frac
    d2-\delta}\Bigg)\!.\nonumber\\
\end{eqnarray}
\end{subequations}

The physical meaning of parameter $\nu^{(+)}$ is clear from
Fig.~\ref{fig:circles}. It represents the kinetic transverse energy
at the last atomic plane before the reflection, divided by the
centrifugal potential difference between the neighboring atomic
planes.

\subsection{Negative particles}\label{sec:neg-part}
In contrast to trigonometric arc-sine, hyperbolic arc-sine function
exists at any value of its argument. Therefore, expression
(\ref{rn}) for negatively charged particle trajectories holds until
the radicand in the motion amplitude $A_n$ given by (\ref{An})
becomes negative. The first interval at which that happens will be
called ``inflection" one. Its order number is inferred to be
\begin{equation}\label{ninfl-def}
    n_{\mathrm{infl}}=\left\lfloor\frac{\tau^2\theta_0^2-(b-\delta)^2}{2\delta
    d}\right\rfloor+1.
\end{equation}
In the inflection interval the amplitude $A_{n_{\mathrm{infl}}}$
calculated by the formula (\ref{An}) would be imaginary. That
implies that the $r(t)$ dependence now is to be described by a
hyperbolic cosine rather than a sine (hence the term ``inflection").
Matching the amplitude and the phase of the hyperbolic cosine with
solution (\ref{rn}) for the preceding $n=n_{\mathrm{infl}}-1$ gives
\begin{eqnarray}\label{r-infl}
    r_{n_{\mathrm{infl}}}(t)\!=\!-\!\delta\!-\!n_{\mathrm{infl}}d\!
    +\!|\!A_{n_{\mathrm{infl}}}\!|\cosh\!\Bigg(\!\frac
    t\tau\!+\!\varphi_0\!+\!\sum_{m=1}^{n_{\mathrm{infl}}-1}\!\!\triangle\varphi_m\nonumber\\
    -\mathrm{arsinh}\frac{\frac d2-\delta}{A_{n_{\mathrm{infl}}-1}}-\mathrm{arcosh}\frac{\frac d2+\delta}{|A_{n_{\mathrm{infl}}}|}\Bigg),\nonumber\\
    \left(-\frac d2-n_{\mathrm{infl}}d\leq r_{n_{\mathrm{infl}}}\leq\frac
    d2-n_{\mathrm{infl}}d\right)\qquad\qquad
\end{eqnarray}
where
\begin{eqnarray}\label{32}
    |A_{n_{\mathrm{infl}}}|&=&\sqrt{-\tau^2\theta_0^2+(b-\delta)^2+2n_{\mathrm{infl}}\delta
    d}\nonumber\\
    &\equiv&\sqrt{2\delta d\left(1-\nu^{(-)}\right)},
\end{eqnarray}
with
\begin{equation}\label{nu-def}
    \nu^{(-)}=\left\{\frac{\tau^2\theta_0^2-(b-\delta)^2}{2\delta
    d}\right\}_\mathrm{f}.
\end{equation}
Since $\frac d2+\delta\geq\sqrt{2\delta d}$, the argument of
$\mathrm{arcosh}$ in (\ref{r-infl}) is always $\geq1$.

Next, a question arises, at which condition the trajectory
(\ref{r-infl}) can actually reach the next interval, i. e.
$r_{n_{\mathrm{infl}}}(t)$ can descend to value
$n_{\mathrm{infl}}d-\frac d2$. Since by Eq.~(\ref{nu-def}),
$r_{n_{\mathrm{infl}}}(t)\geq-\delta-n_{\mathrm{infl}}d+|A_{n_{\mathrm{infl}}}|$,
that would require
\begin{equation}\label{cond-}
    \delta-\frac d2>|A_{n_{\mathrm{infl}}}|.
\end{equation}
Substituting here (\ref{32}), and solving with respect to the ratio
$\frac\delta d$, one may present (\ref{cond-}) in form
\begin{equation}\label{delta>d2}
    \frac\delta d\equiv\frac{R_c}{2R}>f\big(\nu^{(-)}\big)
\end{equation}
with
\begin{equation}\label{}
    f\big(\nu^{(-)}\big)=\frac32-\nu^{(-)}+\sqrt{\left(2-\nu^{(-)}\right)\left(1-\nu^{(-)}\right)}.
\end{equation}
Function $f\big(\nu^{(-)}\big)$ decreases monotonously (almost
linearly) from $f(0)=\frac32+\sqrt2\approx2.9$ to $f(1)=\frac12$.

In the simplest case illustrated in Fig.~\ref{fig:hyperb}, when
condition (\ref{delta>d2}) is violated (e.~g., if $\frac\delta
d<\frac12\leq f$), (\ref{ninfl-def}) must be the last interval
reached by the particle, its order number being
\begin{equation}\label{nmax-delta<d2}
    n^{(-)}_{\max}=n_{\mathrm{infl}}=\left\lfloor\frac{\tau^2\theta_0^2-(b-\delta)^2}{2\delta
    d}\right\rfloor+1\quad\left(\frac\delta d\leq f\big(\nu^{(-)}\big)\!\right).
\end{equation}
Expressing $t_\mathrm{refl}$ from equation
$\dot{r}(t_\mathrm{refl})=0$ then gives
\begin{eqnarray}\label{trefl-neg}
    \frac{t_\mathrm{refl}}\tau=\mathrm{arcosh}\frac{\frac d2+\delta}{|A_{n_{\mathrm{infl}}}|}-\mathrm{arsinh}\frac{b-\delta}{A_0}\qquad\qquad\nonumber\\
    +\sum_{n=0}^{n_{\mathrm{infl}}-1}\mathrm{arsinh}\frac{\frac d2-\delta}{A_n}+\sum_{n=1}^{n_{\mathrm{infl}}-1}\mathrm{arsinh}\frac{\frac d2+\delta}{A_n}.
\end{eqnarray}
Reversal of the summation order here leads to an expression
\begin{eqnarray}\label{trefl-}
    \frac{t_\mathrm{refl}}\tau=\mathrm{arcosh}\frac{\frac d2+\delta}{\sqrt{2\delta d\left(1-\nu^{(-)}\right)}}-\mathrm{arsinh}\frac{b-\delta}{A_0}\qquad\nonumber\\
    +\sum^{n_{\mathrm{infl}}-1}_{n=0}\mathrm{arsinh}\frac{\frac d2-\delta}{\sqrt{2\delta d\left(\nu^{(-)}+n\right)}}\qquad\nonumber\\
    +\sum^{n_{\mathrm{infl}}-2}_{n=0}\mathrm{arsinh}\frac{\frac d2+\delta}{\sqrt{2\delta
    d\left(\nu^{(-)}+n\right)}}.\qquad
\end{eqnarray}
Therethrough, using Eq.~(\ref{thetarefl}), results the deflection
angle.

Otherwise, i. e. if (\ref{delta>d2}) holds (e. g., if $\frac\delta
d>3>f$), in all the subsequent intervals after (\ref{ninfl-def}) the
trajectory must also express through hyperbolic cosines:
\begin{eqnarray}\label{r-beyond}
    r_{n}(t)=-\delta-nd
    +|A_{n}|\cosh\Bigg(\frac
    t\tau+\varphi_0+\!\sum_{m=1}^{n_{\mathrm{infl}}-1}\!\triangle\varphi_m\nonumber\\
    -\mathrm{arsinh}\frac{\frac d2-\delta}{A_{n_{\mathrm{infl}}-1}}-\mathrm{arcosh}\frac{\frac
    d2+\delta}{|A_{n_{\mathrm{infl}}}|}\qquad\qquad\nonumber\\
    -\!\sum_{m=n_{\mathrm{infl}}+1}^{n}\!\mathrm{arcosh}\frac{\frac
    d2+\delta}{|A_m|}+\sum_{m=n_{\mathrm{infl}}}^{n-1}\mathrm{arcosh}\frac{-\frac
    d2+\delta}{|A_m|}\Bigg)\!\quad\\
    \left(-\frac d2-nd\leq r_n\leq\frac
    d2-nd,\qquad n\geq n_{\mathrm{infl}}\right),\qquad\nonumber
\end{eqnarray}
with amplitudes $A_n$ still given by Eq.~(\ref{An}). Sequence
(\ref{r-beyond}) may continue as long as the arguments of all
$\mathrm{arcosh}$ exceed unity, i. e. as long as
\begin{equation}\label{}
    \frac d2+\delta\geq|A_n|
\end{equation}
(which is equivalent to $-\frac d2+\delta\geq|A_{n-1}|$). Inserting
here (\ref{An}), one ultimately infers the value of the reflection
interval order number:
\begin{equation}\label{41}
    n^{(-)}_{\max}=\left\lfloor\frac{\tau^2\theta_0^2-(b-\delta)^2+\left(\frac d2+\delta\right)^2}{2\delta
    d}\right\rfloor\quad
    \left(\frac\delta d>f\big(\nu^{(-)}\big)\!\right).\quad
\end{equation}
The above expression is similar to Eq.~(\ref{nmax+def}) for
positively charged particles. As one might expect, in the
high-energy limit $\delta\gg d,\,b$ values $n^{(+)}_{\max}$ and
$n^{(-)}_{\max}$ coincide and do not depend on the particle energy.

Expressing $t_\mathrm{refl}$ from $\dot{r}(t_\mathrm{refl})=0$ and
Eq.~(\ref{r-beyond}) in this case gives
\begin{subequations}
\begin{eqnarray}\
    \frac{t_\mathrm{refl}}\tau&=&-\mathrm{arsinh}\frac{b-\delta}{A_0}\qquad\qquad\nonumber\\
    &\!&+\sum_{n=0}^{n_{\mathrm{infl}}-1}\mathrm{arsinh}\frac{\frac d2-\delta}{A_n}+\sum_{n=1}^{n_{\mathrm{infl}}-1}\mathrm{arsinh}\frac{\frac
    d2+\delta}{A_n}.\nonumber\\
    &\!&+\sum_{m=n_{\mathrm{infl}}}^{n^{(-)}_{\max}}\mathrm{arcosh}\frac{\frac
    d2+\delta}{|A_m|}\nonumber\\
    &\,&-\sum_{m=n_{\mathrm{infl}}}^{n^{(-)}_{\max}-1}\mathrm{arcosh}\frac{-\frac
    d2+\delta}{|A_m|}\nonumber
    \\&\equiv&-\mathrm{arsinh}\frac{b-\delta}{\sqrt{2\delta d(n_{\mathrm{infl}}-1+\nu^{(-)})}}\qquad\qquad\label{trefl-neg-1}\\
    &\,&+\sum^{n_{\mathrm{infl}}-1}_{n=0}\mathrm{arsinh}\frac{\frac d2-\delta}{\sqrt{2\delta d\left(\nu^{(-)}+n\right)}}\nonumber\\
    &\,&+\sum^{n_{\mathrm{infl}}-2}_{n=0}\mathrm{arsinh}\frac{\frac d2+\delta}{\sqrt{2\delta
    d\left(\nu^{(-)}+n\right)}},\nonumber\\
    &\,&+\sum_{n=1}^{n^{(-)}_{\max}-n_{\mathrm{infl}}+1}\mathrm{arcosh}\frac{\frac d2+\delta}{\sqrt{2\delta
    d\left(n-\nu^{(-)}\right)}}\nonumber\\
    &\,&-\sum_{n=1}^{n^{(-)}_{\max}-n_{\mathrm{infl}}}\mathrm{arcosh}\frac{-\frac d2+\delta}{\sqrt{2\delta
    d\left(n-\nu^{(-)}\right)}}.\label{trefl-neg-2}
\end{eqnarray}
\end{subequations}

Actually, equations (\ref{trefl-neg-2}) can be used not only under
condition (\ref{delta>d2}), but also at any ratio $\frac\delta d$,
provided that in capacity of $n^{(-)}_{\max}$ one uses expression
\begin{equation}\label{nmax-unif}
    n^{(-)}_{\max}=\left\lfloor\frac{\tau^2\theta_0^2-(b-\delta)^2+\left(\frac d2-\delta\right)^2\Theta\left(\frac\delta d
    -\!f\!\big(\nu^{(-)}\!\big)\!\right)}{2\delta
    d}\right\rfloor+1
\end{equation}
(with $\Theta(v)$ the Heavyside unit-step function) unifying
(\ref{nmax-delta<d2}) and (\ref{41}). The universally valid formula
(\ref{nmax-unif}) may be convenient when $t_\mathrm{refl}$ is
evaluated with the aid of computer for widely changing values of
particle energy or crystal bending radius.

As for the physical meaning of $\nu^{(-)}$, at $\frac\delta d<f$,
i.~e. when inflection interval is also that of reflection,
Fig.~\ref{fig:hyperb} illustrates that the meaning of $\nu^{(-)}$ is
similar to that of $\nu^{(+)}$. It is the (appropriately rescaled)
kinetic transverse energy upon the particle entrance to the
reflection inter-planar interval, only the interval boundary now is
not the atomic plane but the last potential maximum passed. In case
if $\frac\delta d>f$, $\nu^{(-)}$ does not characterize the
reflection interval, and vice versa, the kinetic energy in the
reflection interval is not closely related with $\nu^{(-)}$.

The obtained expressions (\ref{rn}-\ref{tn}) for the trajectory and
(\ref{trefl2-c}, \ref{trefl-neg-2}) for its reflection point allow
evaluating all the observables relevant to the particle passage. In
the present paper, we will be interested only in the final angle of
elastic reflection.

\subsection{Thick crystal limit (isolation of volume
effects)}\label{subsec:thick-cryst}

Formulas (\ref{trefl2-c}, \ref{trefl-neg-2}), in principle, contain
dependencies both on volume and on boundary effects. In most
practical cases, the deflecting crystal may be regarded as thick,
whence boundary effects are expected to turn negligible. An increase
of the crystal thickness, or more precisely of the distance between
the crystal boundary and the volume reflection point, may be thought
of as an increase of the particle incidence angle $\theta_0$ (see
Fig.~\ref{fig:passage}). Then, it suffices to consider the limit
\begin{equation}\label{thetavolrefldef}
    \theta_\mathrm{refl}\approx\theta_\mathrm{v.r.}=2\lim_{\theta_0/\theta_c\to\infty}\left(-\theta_0+\delta\frac{t_\mathrm{refl}(\theta_0)}{\tau^2}\right).
\end{equation}
With function (\ref{trefl2-c}), or (\ref{trefl-neg-2}), such a limit
must always be finite: indeed, at large $n_{\max}$ the sum over $n$
grows like the corresponding integral, whose asymptotic behavior
straightforwardly evaluates as
\begin{eqnarray}\label{}
    \frac{t_\mathrm{refl}}\tau\!&\sim&\!\int^{n_{\max}} dn\!\left(\frac{\frac d2+\delta}{\sqrt{2(\nu^{(+)}+n)\delta d}}+\frac{\frac d2-\delta}{\sqrt{2(\nu^{(+)}+n)\delta d}}\right)\nonumber\\
    &\sim&\!\sqrt{\frac{2n_{\max}d}{\delta}}\simeq\theta_0{\frac{\tau}{\delta}}.\nonumber
\end{eqnarray}
This leading asymptotic behavior cancels exactly the first term in
(\ref{thetavolrefldef}), whilst calculation of the finite remainder
requires a more accurate evaluation of the sum, which will be our
task in the next section (in application to the limit $R\gg R_c$).

In general, it must be noted that function
$\theta_\mathrm{v.r.}(\tau,\delta,d,b)$, being a dimensionless
function of 4 dimensional variables, may depend only on their 3
dimensionless ratios -- say, $d/\tau$, $\delta/d$, $b/d$. At that,
the last ratio is always $\sim1$. The first one amounts to
\begin{equation}\label{}
    \frac{d}\tau=\sqrt{\frac{2d}{R_c}}=2\theta_c,
\end{equation}
where $\theta_c$ is the Lindhard critical angle \cite{Lindhard}; so,
it is always small, once we are in a high-energy regime. As for the
ratio
\begin{equation}\label{delta-d}
    \frac{2\delta}d=\frac{R_c}{R},
\end{equation}
it may be either large or small depending on the particle energy and
the crystal bending radius. The regime of particle passage through
the crystal is determined solely by ratio (\ref{delta-d}).

\begin{figure}
\includegraphics{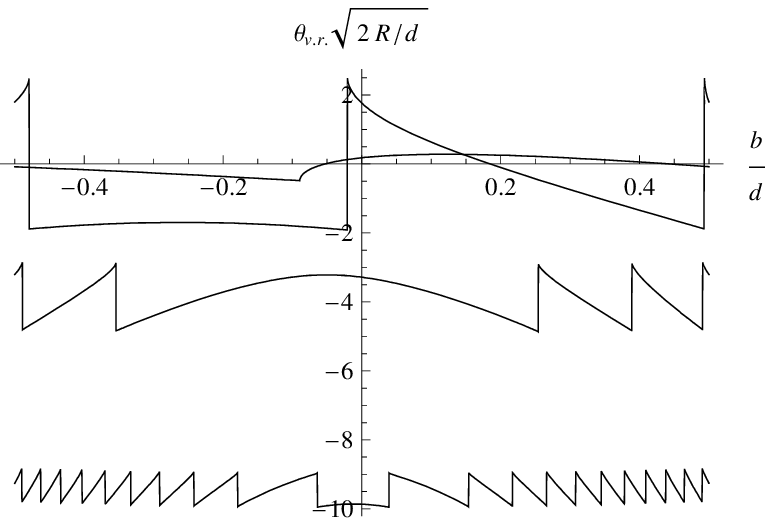}
\includegraphics{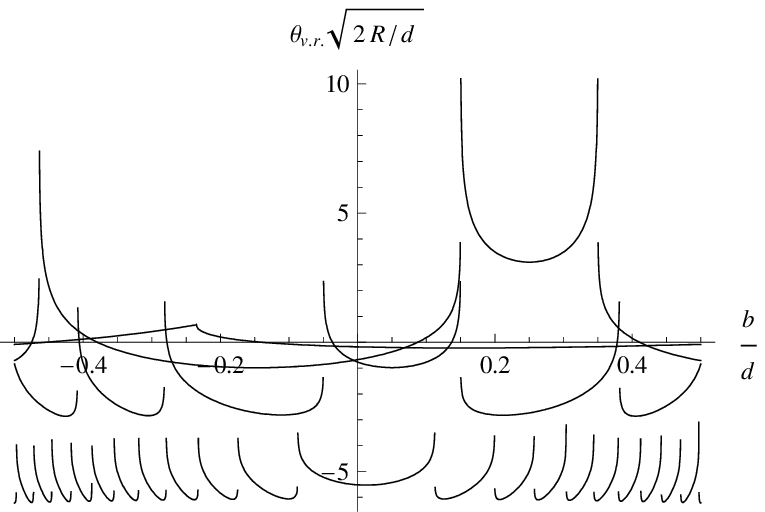}
\caption{\label{fig:theta(b)} The volume reflection angle
$\theta_{\mathrm{v.r.}}$ as a function of impact parameter $b$, for
$R/R_c=1/3, 2, 10, 40$. Top panel -- for positively charged
particles, bottom panel -- for negatively charged particles. Beyond
the shown unit interval the picture repeats periodically.}
\end{figure}

\begin{figure}
\includegraphics{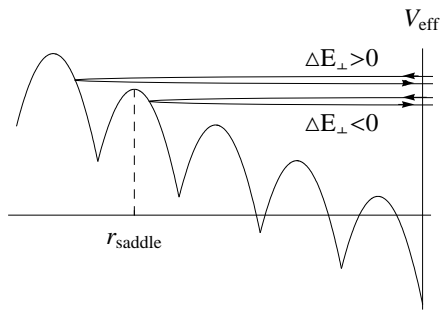}
\caption{\label{fig:orbiting} The relation between the particle
transverse energy and the effective potential energy
$V_{\mathrm{eff}}$ (including the centrifugal potential) under the
conditions of \emph{negatively} charged particle orbiting in a bent
crystal. The leading logarithmic contribution to integral
(\ref{thetarefl-simpler}) comes from the vicinity of point
$r_{\mathrm{saddle}}$ -- the coordinate of the effective potential
maximum to which the particle transverse energy happens to be close.
In the case $\triangle E_\perp>0$ the particle sweeps the two-sided
neighborhood of $r_{\mathrm{saddle}}$, whereas at $\triangle
E_\perp<0$ -- its one-sided neighborhood only.}
\end{figure}

To gain a general impression of functional dependencies involved,
and to test our generic formulas (\ref{trefl2-c},
\ref{trefl-neg-2}), let us view the dependence
$\theta_{\mathrm{v.r.}}(b)$ for different values of $R/R_c$.
Fig.~\ref{fig:theta(b)} shows this dependencies for positive and for
negative particles. They are in fair agreement with Figs.~6 and 8 of
\cite{Maish-arXiv}. But we will pay more attention to interpretation
of the features observed in the figures:
\begin{enumerate}
  \item The origin of the recurrent structure in variable $b$ with a
tapering period is, obviously, due to $\theta_{\mathrm{v.r.}}$
dependence on $b$ through $\nu^{(\pm)}$ alone (see
Eqs.~(\ref{nu+def}), (\ref{nu-def})), insofar as $\nu^{(\pm)}$,
involving an operation of fractional part, is a periodic function of
$\frac{(b\pm\delta)^2}{2\delta d}$, which in the interval $-\frac
d2<b<\frac d2$ makes $\sim2\frac{d^2}{8\delta d}=\frac{R}{2R_c}\gg1$
periods. Also, since $\nu^{(\pm)}$ is an even function of
$b\pm\delta$, the particle deflection angle is a symmetric function
of $b$ with respect to point $b=-\delta$ for positively charged
particles, and with respect to $b=\delta$ for negative particles.
  \item Another feature of $\theta_{\mathrm{v.r.}}(b)$ dependencie(s) is
that, for negatively charged particles, the reflection angle blows
up (formally) to $+\infty$ at certain values of impact parameters.
Physically, that corresponds to close matching of the particle
transverse energy to the height of a (locally parabolic) effective
potential barrier -- the situation known as orbiting (see
\cite{rainbow}) \footnote{The verbal description of the given effect
for negative particle case is contained in \cite{Tar-Vor} (end of
Sec.~3), and a graphical illustration thereof appears in
\cite{Maisheev} (Fig.~5, trajectory 2). But relation with the
general concept of orbiting (spiral scattering), as formulated in
\cite{rainbow}, was only noted in \cite{Kovalev}. We don't go into
discussion of possible experimental significance of orbiting here.}.
The asymptotics of the divergences is logarithmic \cite{rainbow}, as
follows from the general integral expression of the deflection angle
in a central potential $V(r)$ \footnote{Equation (\ref{theta-cs})
agrees with our definition of the $\theta$ angle and differs in sign
from the conventionally defined angle in a centrally-symmetric
field.}:
\begin{subequations}
\begin{eqnarray}
  \theta &\underset{E\gg m}{\approx}& 2M\int^\infty_{r_{\min}}\frac{dr/(R+r)^2}{\!\sqrt{\left(E\!-\!V(r)\right)^2\!-\!M^2\!/(R\!+\!r)^2}}-\pi \label{theta-cs}\\
  &\underset{R\gg r,b}{\approx}&\frac1R\int^\infty_{r_{\min}}\frac{dr}{\sqrt{\frac{\theta_0^2}4+\frac{V_{\mathrm{eff}}(b)-V_{\mathrm{eff}}(r)}{2E}}}-\pi\label{thetarefl-simpler}\\
  &\underset{\mathrm{orbit.}}\sim& \theta_c\frac{4\delta}{d}\!\!\underset{|r-r_{\mathrm{saddle}}|\lesssim d}\int\frac{dr}{\sqrt{\frac{\triangle E_\perp}{\left|F_{\max}\right|}\!+\!(r\!-\!r_{\mathrm{saddle}})^2}}\!+\ldots\qquad\label{log-int}\\
  &\simeq&\!\!\!\theta_c\frac{4\delta}{d}\Bigg\{\begin{array}{c}
                            \ln\frac{1}{\nu^{(-)}}+...\qquad (\triangle E_\perp\propto\nu^{(-)}\to+0)\qquad\,\\
                            \frac12\ln\frac{1}{1-\nu^{(-)}}+...\,\,(\triangle E_\perp\propto\nu^{(-)}-1\to-0).\\
  \end{array}\nonumber\\
  \label{log-log}
\end{eqnarray}
\end{subequations}
Here
\[
M\simeq(R-b)\left[E-V(b)\right]\left(1-\theta_0^2/2\right)
\]
is the particle angular momentum relative to the crystal bend axis,
\begin{equation}\label{V-eff}
    V_{\mathrm{eff}}(r)=V(r)-E\frac rR
\end{equation}
is the effective potential including the centrifugal energy,
$r_{\mathrm{saddle}}$ -- the position of maximum of the effective
potential barrier whose height in the case of orbiting happens to be
close to the particle energy, and $\triangle E_\perp$ -- the
transverse energy variation relative to the height of the effective
potential barrier. The factor $\frac12$ in the $\triangle E_\perp<0$
alternative of Eq.~(\ref{log-log}) arises because the integration in
(\ref{log-int}) is then carried out only over the one-sided
neighborhood of $r_{\mathrm{saddle}}$ where the radicand stays
positive (see Fig.~\ref{fig:orbiting}).
  \item It must be noticed that for negative particles function $\theta_{\mathrm{v.r.}}(b)$
has smooth minima, which must correspond to caustics, i. e., to
rainbow scattering \cite{rainbow}.
  \item In contrast, for positive particles the potential in its maximum is
not differentiable, excluding both orbiting and rainbow scattering
for this case. With some smearing of the potential around the atomic
planes, these affects, of course, re-appear.
\end{enumerate}

The proper question is whether it is possible to derive at least the
particle final deflection angle (related with
$\frac{t_{\mathrm{refl}}}\tau$) from the more conventional integral
representation approach \cite{Tar-Vor,Maisheev}. In that approach
momentum and transverse energy conservation laws are incorporated
automatically, so there is no need to connect trajectories on the
interval borders. Indeed, specializing in (\ref{V-eff},
\ref{thetarefl-simpler})
\begin{eqnarray}\label{}
    V(r)=F_{\mathrm{max}}d\left(\frac
    rd+n\right)^2,\quad n=-\left\lfloor\frac
    rd+\frac12\right\rfloor\\
    \left(\mathrm{at} \,\,r<0\,\, n\geq0\right)\quad\qquad\qquad\nonumber
\end{eqnarray}
and using basic integral
\begin{eqnarray}
  \int^{d/2-nd} _{-d/2-nd}\frac{dr}{\sqrt{\tau^2\theta^2_0+b^2+2\delta(b+r)-(r+nd)^2}} \nonumber\\
   =\arcsin\frac{\frac d2-\delta}{\sqrt{\tau^2\theta^2_0+(b+\delta)^2-2n\delta
   d}} \nonumber\\
   +\arcsin\frac{\frac d2+\delta}{\sqrt{\tau^2\theta^2_0+(b+\delta)^2-2n\delta d}}
\end{eqnarray}
for positively charged particles, and a similar one for negative
particles, we reproduce the inverse trigonometric and hyperbolic
functions encountered in (\ref{trefl}, \ref{trefl-},
\ref{trefl-neg-1}). But the integral representation approach
wouldn't give us explicit trajectories $r(t)$ (rather, $t(r)$, to be
solved for $r$), and the geometric interpretation
(Figs.~\ref{fig:circles}, \ref{fig:hyperb}).

On the other hand, from the integral representation for the final
angle we might derive the result in form of a sum of analytic
functions also for a more complicated parametrization of the
inter-planar potential -- e.~g., adding thereto a term proportional
to $r^4$. Then instead of arcsines one would get elliptic functions.
But it is the simplicity of functions in the sum that permits us, in
the important limit $R\gg R_c$, when the number of terms in the sum
gets large, to replace the sums by integrals and do them in closed
form. In this sense, analytic investigation only \emph{begins} here.

\section{Volume reflection regime (moderately high energies, $R\gg R_c$)}\label{sec:largeR}

As we had mentioned in the Introduction, and as
Fig.~\ref{fig:theta(b)} does confirm, under the condition $R\gg
R_c$, i. e. $2\delta\ll d$, the  particle deflection angles depends
weakly on the impact parameter. So, it is interesting, in the first
place, to determine the numerical value of the limiting ratio
$\lim_{R/R_c\to\infty}\frac{\theta_{\mathrm{v.r.}}}{\theta_c}$.
Secondly, it is desirable to determine the final beam shape and
quantify its angular width as a function of $R/R_c$. That will be
our aim for the present section. The treatment is somewhat different
for the cases of positively and negatively charged particles,
because of the difference between the functional form of initial
Eqs.~(\ref{trefl2-c}) and (\ref{trefl-neg-2}).

\subsection{Positive particles}\label{sec:pos-part}
For positively charged particles, in the considered limiting case
$\delta\ll d$, say, quantity $\left(\frac
d2\pm\delta\right)^2+2\left(\nu^{(+)}+n\right)\delta d$ entering the
denominators in (\ref{trefl2}) varies relatively little as $n$
changes from $n$ to $n+1$. Thus it must be possible to replace the
summation in (\ref{trefl2-c}) by integration. The proper
mathematical tool for that is the Euler-Maclaurin formula (see, e.
g., \cite{Eul-Macl}) which reads
\begin{equation}\label{Eul-Macl}
    \sum_{n=0}^Nf(n)=\frac12f(0)+\int_0^Ndnf(n)+\frac12f(N)+\mathcal{O}\left(\frac{df}{dn}\right).
\end{equation}
Employing this formula for approximation of each of the sums in
(\ref{trefl3}) (which is somewhat more convenient than
Eq.~(\ref{trefl2})), one gets \footnote{In this subsection the sign
alternative in $\frac d2\pm\delta$ corresponds to dealing with the
first or with the second of the sums in (\ref{trefl3}).}
\begin{eqnarray}\label{sum-int+}
    \sum_{n=0}^{n_{\max}^{(+)}-1}\mathrm{arccot}\frac{{\sqrt{2\delta d\left(\nu^{(+)}+n\right)}}}{\frac d2\pm\delta}\qquad\qquad\qquad\qquad\qquad\qquad\qquad\nonumber\\
    =\frac12\mathrm{arccot}\frac{\sqrt{2\delta d\nu^{(+)}}}{\frac d2\pm\delta}+\int^{n_{\max}^{(+)}+\nu^{(+)}-1}_{\nu^{(+)}}dn\,\mathrm{arccot}\frac{\sqrt{2\delta dn}}{\frac d2\pm\delta}\quad\qquad\nonumber\\
        +\frac12\mathrm{arccot}\frac{\sqrt{2\delta d\left(\!n_{\max}^{(+)}\!+\!\nu^{(+)}\!-\!1\!\right)}}{\frac
    d2\pm\delta}\nonumber
    +\mathcal{O}\!\left(\!\!\sqrt{\frac{\delta}{d}}\right),\quad\qquad\qquad\qquad
\end{eqnarray}
where we had estimated, for all $n$,
\begin{equation}\label{}
    \left|\frac d{dn}\mathrm{arccot}\frac{\sqrt{2\delta d\left(\nu^{(+)}+n\right)}}{\frac
    d2\pm\delta}\right|\lesssim\sqrt{\frac{\delta}{d}}.
\end{equation}
The two end-point contributions in the third line of
(\ref{sum-int+}) are small as $\mathcal{O}\left(\delta/d\right)$
relative to the integral, but still they need to be kept if we wish
to describe not only the mean deflection, but also the scattered
beam shape.

Taking the indefinite integral in Eq.~(\ref{sum-int+}) by parts
\[
    \int\!dn\,
    \mathrm{arccot}\sqrt{an}=\frac1a\left[(1\!+\!an)\mathrm{arccot}\sqrt{an}+\sqrt{an}\right],
\]
one brings (\ref{sum-int+}) to the form
\begin{eqnarray}\label{uu}
    \sum_{n=0}^{n_{\max}^{(+)}-1}\mathrm{arccot}\frac{\sqrt{2\delta d(\nu^{(+)}+n)}}{\frac d2\pm\delta}\quad\qquad\qquad\qquad\qquad\qquad\nonumber\\
    \underset{\delta\ll d}\approx\frac{\left(\frac d2\pm\delta\right)^2}{2\delta d}\Bigg(\!\left[1+\frac{2\delta d}{\left(\frac d2\pm\delta\right)^2}\left(n_{\max}^{(+)}+\nu^{(+)}-1\right)\right]\qquad\nonumber\\
    \cdot\mathrm{arccot}\frac{\sqrt{2\delta d\left(n_{\max}^{(+)}+\nu^{(+)}-1\right)}}{\frac d2\pm\delta}\nonumber\\
    +\frac{\sqrt{2\delta d\!\left(n_{\max}^{(+)}\!+\!\nu^{(+)}\!-\!1\right)}}{\frac d2\pm\delta}\,\,\,\qquad\qquad\qquad\qquad\qquad\nonumber\\
    -\left[1+\frac{2\delta d\nu^{(+)}}{\left(\frac d2\pm\delta\right)^2}\right]\mathrm{arccot}\frac{\sqrt{2\delta d\nu^{(+)}}}{\frac d2\pm\delta}
    -\frac{\sqrt{2\delta d\nu^{(+)}}}{\frac d2\pm\delta}\Bigg)\nonumber\\
    +\frac12\mathrm{arccot}\frac{\sqrt{2\delta d\nu^{(+)}}}{\frac d2\pm\delta}\!+\!\frac12\mathrm{arccot}\frac{\sqrt{2\delta d\!\left(\!n_{\max}^{(+)}\!+\!\nu^{(+)}\!-\!1\!\right)}}{\frac d2\pm\delta}\nonumber\\
    +\mathcal{O}\left(\!\!\sqrt{\frac{\delta}{d}}\right).\qquad
\end{eqnarray}
In the limit $n_{\max}^{(+)}\to\infty$, with the use of asymptotic
expansion
$\mathrm{arccot}{\sqrt{\eta}}=\frac\pi2-\sqrt\eta+\mathcal{O}\left(\eta^{3/2}\right)$,
expression (\ref{uu}) reduces to
\begin{eqnarray}\label{nmaxtoinf}
  \sum_{n=0}^{n_{\max}^{(+)}-1}\mathrm{arccot}\frac{\sqrt{2\delta d\left(\nu^{(+)}+n\right)}}{\frac d2\pm\delta}\qquad\qquad\qquad\qquad\qquad\nonumber\\
    \underset{n_{\max}^{(+)}\gg 1}\to\frac{d\pm2\delta}{\sqrt{2\delta d}}\sqrt{n_{\max}^{(+)}}\,\,\qquad\qquad\qquad\qquad\qquad\qquad\qquad\nonumber\\
    \!-\frac{\left(\frac d2\pm\delta\right)^2}{2\delta
    d}\!\Bigg[\!\!\left(\!1\!+\!\!\frac{2\delta
    d}{\left(\frac d2\pm\delta\right)^2}\!\right)\!\!\!\left(\!\frac\pi2\!-\!\frac{\!\!\sqrt{2\delta
    d\nu^{(+)}}}{\frac d2\pm\delta}\!\right)
    \!+\!\frac{\!\sqrt{2\delta
    d\nu^{(+)}}}{\frac d2\pm\delta}\Bigg]\nonumber\\
    +\frac12\cdot\frac\pi2+\mathcal{O}\left(\frac1{\sqrt{n_{\max}^{(+)}}},\sqrt{\frac\delta
    d}\right).\qquad\qquad\qquad\qquad\qquad
\end{eqnarray}
Here, one notices that terms $-\frac{\!\sqrt{2\delta
    d\nu^{(+)}}}{\frac d2\pm\delta}$, $+\frac{\!\sqrt{2\delta
    d\nu^{(+)}}}{\frac d2\pm\delta}$ in the brackets in (\ref{nmaxtoinf}) cancel. Further on, inserting
(\ref{nmaxtoinf}) to (\ref{trefl3}) and this to
(\ref{thetavolrefldef}), we witness the anticipated cancelation of
the large terms $-\theta_0+\frac{\sqrt{2\delta
d}}{\tau}n_{\max}^{(+)}\cong0$, and ultimately arrive at result
\begin{eqnarray}
    \theta_\mathrm{v.r.}(b)\!&\approx&\!-2\theta_0+2\frac\delta\tau\Bigg(\frac\pi2+\frac{2d}{\sqrt{2\delta d}}\sqrt{n_{\max}^{(+)}}\nonumber\\
    &\,&-\frac\pi2\!\left[\frac{\left(\frac d2+\delta\right)^2}{2\delta d}\!+\!\nu^{(+)}\right]\!+\!\frac\pi4\nonumber\\
    &\,&-\frac\pi2\!\left[\frac{\left(\frac d2-\delta\right)^2}{2\delta
    d}\!+\!\nu^{(+)}\right]\!+\!\frac\pi4\Bigg)\nonumber\\
    &\equiv&\!\!-\frac\pi2\theta_c\!\left[1-\frac{4R_c}R\!\left(1-\nu^{(+)}(b)\right)+\!\mathcal{O}\!\left(\frac{R_c^{3/2}}{R^{3/2}}\right)\!\right]\qquad\label{appr+}
\end{eqnarray}
(with $\nu^{(+)}(b)$ being given by Eq.~(\ref{nu+def})).

\begin{figure}
\includegraphics{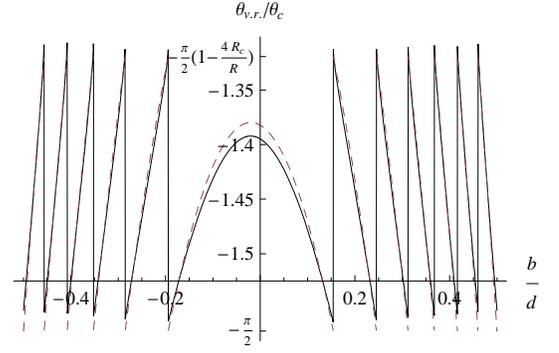}
\caption{\label{fig:approx+} Solid curve -- positively charged
particle reflection angle $\theta_{\mathrm{v.r.}}$ versus impact
parameter $b$, for $R/R_c=25$. Dashed curve -- approximation
(\ref{appr+}). The central segment of the curves is strongly
$\theta_0$-dependent.}
\end{figure}

Comparison of approximation (\ref{appr+}) with the exact result
(\ref{trefl2-c}) is shown in Fig.~\ref{fig:approx+}. (Actually, the
given approximation appears to be numerically accurate starting from
$R/R_c\sim5$). From the figure (or Eq.~(\ref{appr+})) one concludes
that in the first approximation all the particles are deflected to
the same angle $\approx-\frac\pi2\theta_c$. There is also some
dispersal of the scattering angles, depending on the particle impact
parameter, of the full width
\begin{equation}\label{width+}
    \triangle\theta_{\mathrm{v.r.}}=\theta_c\frac{2\pi R_c}{R}\equiv\frac{2\pi\delta}{\tau}.\qquad\mathrm{(posit.\,charged\,particles)}
\end{equation}
The observable quantity, however, is not the indicatrix
$\theta_{\mathrm{v.r.}}(b)$ but the scattering differential
cross-section (final particle flux averaged over the impact
parameters $b$) as a function of the scattering angle
$\theta_{\mathrm{v.r.}}$. Therefore, it is desirable to reconstruct
the latter dependence issuing from the first. That does not pose any
principal problem, granted the linearity of dependence
$\theta_{\mathrm{v.r.}}(\nu^{(+)})$.

\subsubsection{Differential cross-section}

Turning to evaluation of the differential cross-section, one
encounters a certain complication: the $b$-dependent quantity
$\nu^{(+)}$ in (\ref{appr+}) also contains dependence on
$\theta_{0}$.  The fact of residual $\theta_{0}$-dependence was
noticed in \cite{Maish-arXiv}. To some degree, it conflicts with our
initial assumption about the boundary condition vanishing influence
in the limit of large $\theta_0/\theta_c$. We can not revoke it at
the present stage, since in equation (\ref{thetavolrefldef}) we had
already incorporated the facilitating assumption of the trajectory
symmetry with respect to point $t_{\mathrm{refl}}$. Obviously, the
sensitivity to the boundary conditions in general destroys such a
symmetry. Furthermore, $\theta_{\mathrm{refl}}$ might as well
contain a dependence on the particle \emph{exit angle} relative to
the atomic planes, which we did not even take trouble to specify.
Altogether, that may rise a suspicion that the $b$-dependent
correction obtained in (\ref{appr+}) is unreliable for evaluating
the differential cross-section. Fortunately, the impediment is not
fatal and can be overcome within the present framework. In
principle, the differential cross-section sensitivity to $\theta_0$
is attenuated with the increase of $R/R_c$, but more importantly, we
will prove that upon averaging over a \emph{tiny} interval of
$\theta_0$ this dependence is eliminated completely.

To begin with, the differential cross-section involves only a
\emph{derivative} of function $\theta_{\mathrm{v.r.}}(b)$:
\begin{eqnarray}\label{dlambda dtheta}
    \frac{d\lambda}{d\theta_{\mathrm{v.r.}}}&=&\sum_m\frac1{\left|d\theta_{\mathrm{v.r.}}/db\right|_{b=b_m\left(\theta_{\mathrm{v.r.}}\right)}}\nonumber\\
    &\equiv&\frac{R}{2\pi\theta_c
    R_c}\sum_m\frac1{\left|d\nu^{(+)}/db\right|_{b=b_m(\nu^{(+)}(\theta_{\mathrm{v.r.}}))}},\quad
\end{eqnarray}
where $b_m(\theta)$ is the set of all the roots of equation
$\theta_{\mathrm{v.r.}}(b)=\theta$ belonging to the interval $-\frac
d2<b<\frac d2$. Now, at $R/R_c\gg1$ the number of roots $b_m$ of
equation $\theta_{\mathrm{v.r.}}=\theta$ is large, and so, in
general, they are densely distributed over the finite definition
interval $-\frac d2<b<\frac d2$. It appears that the root
distribution density is just proportional to the derivative in the
denominator of (\ref{dlambda dtheta}) (the formal demonstration is
relegated to the Appendix). Therefore, the sum appearing in
(\ref{dlambda dtheta}) approximately equals to just the $b$
variation interval length, i. e., $d$. However, the relation
expected thereby,
\begin{equation}\label{plateau}
    \frac{d\lambda}{d\theta_{\mathrm{v.r.}}}\simeq \frac{Rd}{2\pi\theta_c
    R_c},
\end{equation}
does not yet hold \emph{uniformly} in $b$, and hence in
$\theta_{\mathrm{v.r.}}$. For instance, in a neighborhood of point
$b=-\delta$ we have in the denominator of (\ref{dlambda dtheta})
$\partial\nu^{(+)}/\partial b\to0$ (see Fig.~\ref{fig:approx+}), so
there the differential cross-section blows up above the plateau
(\ref{plateau}) (see Appendix). But the latter peak position on the
$\theta_{\mathrm{v.r.}}$ axis depends sharply on the value of
$\theta_0$ and hence is essentially ``random", needing to be
averaged over.

\begin{figure}
\includegraphics{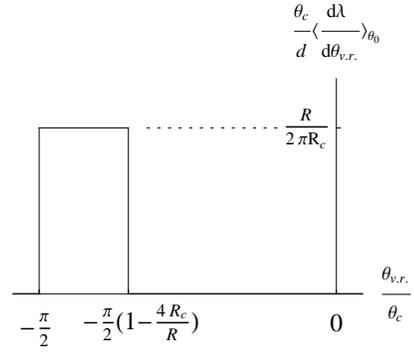}
\caption{\label{fig:diff-cross+} Asymptotic (at $R\gg R_c$) behavior
of the $\theta_0$-averaged differential cross-section for positively
charged particle scattering (Eq.~(\ref{diff-cross+aver})). The area
under the rectangular curve is unity, representing the total
probability. In higher orders in $R_c/R$ the distribution edges must
smear out (see discussion in the text).}
\end{figure}

Indeed, one notices that the dependence of $\nu^{(+)}$ on $\theta_0$
is quadratic, so a situation is possible when the incident particle
beam divergence is smaller than the angular spread acquired in the
crystal:
\begin{equation}\label{11}
    \triangle\theta_0\ll \triangle\theta_{\mathrm{v.r.}},
\end{equation}
but at the same time, the indeterminance of
$\frac{\tau^2\theta_0^2}{2\delta d}\approx n^{(+)}_{\max}$ is
greater than unity:
\begin{equation}\label{22}
    \triangle\left(\frac{\tau^2\theta_0^2}{2\delta d}\right)=\frac{\tau^2\theta_0}{\delta
    d}\triangle\theta_0\gg1.
\end{equation}
Together, Eqs.~(\ref{11}, \ref{22}) may be viewed as a double
inequality
\begin{equation}\label{ineqs}
    \frac{2\delta}{\tau}\frac{\theta_c}{\theta_0}\ll\triangle\theta_0\ll\frac{2\pi\delta}{\tau}\qquad(\theta_0\mathrm{-averaging}).
\end{equation}
Here the sufficient gap exists provided
\begin{equation}\label{condition}
    \theta_0\ggg\frac{\theta_c}\pi.
\end{equation}
This is basically the same condition that we had assumed in writing
Eq.~(\ref{thetavolrefldef}), thus for derivation of a
$\theta_0$-averaged differential cross-section we can safely rely on
Eq.~(\ref{appr+}).

Ultimately, we can make a specific statement that under conditions
(\ref{ineqs}), upon $\theta_0$-averaging, the differential
cross-section equals to constant (\ref{plateau}) over an interval
where roots $b_m$ exist. This interval is determined in the Appendix
(Eq.~({theta-interval})). So, the $\theta_0$-averaged differential
cross-section (the final beam profile) is described by a simple
rectangular function
\begin{eqnarray}\label{diff-cross+aver}
    \left\langle\frac{d\lambda}{d\theta_{\mathrm{v.r.}}}\right\rangle_{\theta_0}\approx\frac{Rd}{2\pi\theta_cR_c}\quad\qquad\qquad\qquad\qquad\qquad\qquad\nonumber\\
    \cdot\Theta\!\left(\theta_{\mathrm{v.r.}}\!+\frac\pi2\theta_c\right)\Theta\!\left(\!-\theta_{\mathrm{v.r.}}-\frac\pi2\theta_c\!\left(1-\frac{4R_c}{R}\right)\!\right)\quad
\end{eqnarray}
(see Fig.~\ref{fig:diff-cross+}).

\subsubsection{Comparison with experiment}

The deflection angle mean value is least affected by multiple
scattering, and thus, may be directly compared with the experiment.
From (\ref{appr+}) we obviously infer
\begin{equation}\label{lin}
    \left\langle\theta_{\mathrm{v.r.}}\right\rangle=-\frac\pi2\theta_c\left(1-\frac{2R_c}R\right)
    \equiv-\frac\pi2\sqrt{\frac d{2R_c}}\left(1-\frac{2R_c}R\right).
\end{equation}
The property of (\ref{lin}) is the linearity of dependence on the
crystal curvature $R^{-1}$; the linear kind of dependence was indeed
noticed in CERN experiments with
$E=400\,\mathrm{GeV}\,\mathrm{GeV/cm}$ \cite{Scand-linear}.

\begin{figure}
\includegraphics{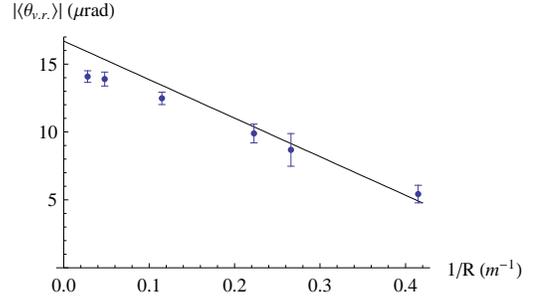}
\caption{\label{fig:fit} Mean volume reflection angle dependence on
the crystal curvature, at $E=400\,\mathrm{GeV}$, for silicon crystal
in orientation (110). Points -- experimental data from
\cite{Scand-linear}. The line -- prediction of Eq.~(\ref{lin}), with
parameters evaluated as explained in the text.}
\end{figure}

To make quantitative comparison with the experiment, one needs to
specify the potential strength in our model. In reality, the Si
(110) inter-planar potential is characterized by 2 parameters:
$F_{\max}\approx6\,\mathrm{GeV/cm}$ (usually used for evaluation of
$R_c$ for channeling processes) and the well depth
$V_0=22.7\,\mathrm{eV}$ (usually used for evaluation of the critical
angle $\theta_c$ for volume reflection). The relation
$V_0=\frac14F_{\max}d$ implied by a quadratic potential model only
holds with accuracy $\approx20\%$:
\begin{eqnarray*}
22.7\,\mathrm{eV}=V_0\neq\frac14F_{\max}d=28.8\,\mathrm{eV}\nonumber\\
(d=1.92\AA \,\,\,\mathrm{for\,\,\,Si\,\,(110)}).\qquad
\end{eqnarray*}
If we evaluate $R_c$ in (\ref{lin}) as
$R_c=E/F_{\max}=0.67\,\mathrm{m}$, it will produce too large
$\left|\left\langle\theta_{\mathrm{v.r.}}\right\rangle\right|$. But
evaluating both $R_c$ and $\theta_c$ as
$R_c=\frac{Ed}{4V_0}\approx0.85 \,\mathrm{m}$,
$\theta_c=\sqrt{2V_0/E}\approx11\,\mu\mathrm{rad}$, and substituting
to Eq.~(\ref{lin}), we get a satisfactory agreement with the
experiment (see Fig.~\ref{fig:fit}).

As for the obtained rectangular profile shape, it is more sensitive
to multiple scattering, and was not yet probed by experiments (the
optimal experimental conditions will be specified in
Sec.~\ref{sec:optimiz}). But we can compare our profile with the
available numerical simulation results using a realistic, smeared
potential, without multiple scattering: \cite{Maish-arXiv}, Fig.~6.
In that case, the positive particle profile shows indeed a signature
of flattening (``shoulder") but near its edges the distribution
behaves differently, exhibiting a subtle divergence (rainbow) at the
outer edge, and decreasing continuously on the inward side. So, for
positively charged particles our simplified model of parabolic
inter-planar potential describes the final beam profile only
quantitatively, though it is able to predict the distribution width
and mean value.

Next, turning to the negative particle reflection problem, we shall
see that in this case our analytic approach is able to capture also
the final beam profile edge details.

\subsection{Negative particles}
In the case of negative particles, the starting point is
Eq.~(\ref{trefl-}) (relevant under (strong) condition
(\ref{delta>d2})), and again, it has to be examined with the object
to trade the sum for an integral. First of all, it has to be minded
that at $n\sim1$ the hyperbolic arcsine arguments vary
significantly, but at the same time they are large, whereas
hyperbolic arcsine of a large argument is close to logarithm of a
large argument: $\mathrm{arsinh}\,v\underset{v\gg1}\simeq\ln2v$, and
thus varies relatively little. On the other hand, in the domain of
large $n$ the arguments of the arcsines vary little. Therefore, over
the \emph{entire} summation interval both sums involved may be
approximated via integrals. Yet, first terms in the sums are
singular functions of $\nu^{(-)}$, and therefore are better taken
into account separately. Thereby, application of the Euler-Maclaurin
formula to the first of the sums in Eq.~(\ref{trefl-}) gives
\begin{eqnarray}\label{sum-int-}
  \sum^{n_{\mathrm{infl}}-1}_{n=0}\mathrm{arsinh}\frac{\frac d2-\delta}{\sqrt{2\delta d\left(\nu^{(-)}+n\right)}}\qquad\qquad\nonumber\\
    \approx\ln\frac{d-2\delta}{\sqrt{2\delta d\nu^{(-)}}}+\frac12\ln\frac{d-2\delta}{\sqrt{2\delta
    d\left(\nu^{(-)}+1\right)}}\nonumber\\
    +\int_{\nu^{(-)}+1}^{\nu^{(-)}+n_{\mathrm{infl}}-1}\!dn~\mathrm{arsinh}\frac{\frac d2-\delta}{\sqrt{2\delta d}\sqrt{n}}\qquad\nonumber\\
    +\frac12\mathrm{arsinh}\frac{\frac d2-\delta}{\sqrt{2\delta
    d\left(\nu^{(-)}+n_{\mathrm{infl}}-1\right)}}.\qquad
\end{eqnarray}
The next-to-leading order (derivative-related) correction term
\cite{Eul-Macl} to (\ref{sum-int-}) amounts
\begin{equation}\label{}
    \frac1{12}\frac{d}{dn}\mathrm{arsinh}\frac{\frac d2-\delta}{\sqrt{2\delta
    d\!\left(\nu^{(-)}\!+\!n\right)}}\Bigg|_{n=1}^{n=n_{\mathrm{infl}}-1}\approx-\frac1{24\left(1\!+\!\nu^{(-)}\right)}.
\end{equation}
We will omit it because of the smallness of the numerical
coefficient $\frac1{24}$, although, in principle, asymptotically it
is also relevant (the same is true for all the higher derivatives,
whose contributions enter with yet smaller coefficients (involving
inverse factorial and Bernoulli numbers)).

Next, calculating the indefinite integral in Eq.~(\ref{sum-int-}) by
parts,
\begin{equation}\label{}
    \int
    dn\,\mathrm{arsinh}\frac{a}{\sqrt n}=n\,\mathrm{arsinh}\frac{a}{\sqrt
    n}+a\sqrt{a^2+n},
\end{equation}
we bring (\ref{sum-int-}) to form
\begin{eqnarray}\label{asum}
  \sum^{n_{\mathrm{infl}}-1}_{n=0}\mathrm{arsinh}\frac{\frac d2-\delta}{\sqrt{2\delta d}\sqrt{\nu^{(-)}+n}}\quad\qquad\qquad\qquad\qquad\qquad\nonumber\\
    \approx\left(\nu^{(-)}\!+\!n_{\mathrm{infl}}\!-\!1\right)\mathrm{arsinh}\frac{\frac d2-\delta}{\sqrt{2\delta d\left(\nu^{(-)}\!+\!n_{\mathrm{infl}}\!-\!1\right)}}\quad\nonumber\\
    +\frac{\frac d2-\delta}{\sqrt{2\delta d}}\sqrt{\frac{\left(\frac d2-\delta\right)^2}{2\delta d}+\nu^{(-)}+n_{\mathrm{infl}}-1}\quad\qquad\qquad\qquad\nonumber\\
    -\left(1+\nu^{(-)}\right)\mathrm{arsinh}\frac{\frac d2-\delta}{\sqrt{2\delta d\left(1+\nu^{(-)}\right)}}\quad\qquad\qquad\qquad\nonumber\\
    -\frac{\frac d2-\delta}{\sqrt{2\delta d}}\sqrt{\frac{\left(\frac d2-\delta\right)^2}{2\delta d}+1+\nu^{(-)}}+\ln\frac{d-2\delta}{\sqrt{2\delta d\nu^{(-)}}}\qquad\nonumber\\
    +\frac12\!\ln\!\frac{d-2\delta}{\!\sqrt{2\delta
    d\!\left(1\!+\!\nu^{(-)}\right)\!}}
    +\frac12\mathrm{arsinh}\frac{\frac d2-\delta}{\!\sqrt{2\delta
    d\!\left(\nu^{(-)}\!+\!n_{\mathrm{infl}}\!-\!1\right)\!}}.\,\,\nonumber\\
\end{eqnarray}

Now, in the thick-crystal limit $n_{\mathrm{infl}}\to\infty$,
Eq.~(\ref{asum}) simplifies to
\begin{eqnarray}\label{}
  \sum^{n_{\mathrm{infl}}-1}_{n=0}\mathrm{arsinh}\frac{\frac d2-\delta}{\sqrt{2\delta d}\sqrt{\nu^{(-)}+n}}\quad\qquad\qquad\qquad\qquad\qquad\nonumber\\
    \underset{n_{\mathrm{infl}}\gg1}\to\sqrt{n_{\mathrm{infl}}}\frac{d-2\delta}{\sqrt{2\delta d}}
    -\left(1+\nu^{(-)}\right)\ln\frac{d-2\delta}{\sqrt{2\delta d\left(1+\nu^{(-)}\right)}}\nonumber\\
    -\frac{\left(\frac d2-\delta\right)^2}{2\delta d}\sqrt{1+\frac{2\delta d}{\left(\frac d2-\delta\right)^2}\left(1+\nu^{(-)}\right)}+\ln\frac{d-2\delta}{\sqrt{2\delta d\nu^{(-)}}}\nonumber\\
    +\frac12\ln\frac{d-2\delta}{\!\sqrt{2\delta
    d\left(1+\nu^{(-)}\right)}}\qquad\qquad\qquad\qquad\qquad\qquad\nonumber\\
  \cong\sqrt{n_{\mathrm{infl}}}\frac{d-2\delta}{\sqrt{2\delta d}}
    -\left(\frac12+\nu^{(-)}\right)\ln\frac{d-2\delta}{\sqrt{2\delta d\left(1+\nu^{(-)}\right)}}\quad\nonumber\\
    -\frac{\left(\frac d2-\delta\right)^2}{2\delta d}
    -\frac12\big(1+\nu^{(-)}\big)+\ln\frac{d-2\delta}{\sqrt{2\delta
    d\nu^{(-)}}}.\qquad\qquad\nonumber\\
    \label{sum1}
\end{eqnarray}
Similarly, the second term in (\ref{trefl-}) reduces to
\begin{eqnarray}
  \sum^{n_{\mathrm{infl}}-2}_{n=0}\mathrm{arsinh}\frac{\frac d2+\delta}{\sqrt{2\delta d\left(\nu^{(-)}+n\right)}}\qquad\qquad\qquad\qquad\qquad\nonumber\\
    \to\sqrt{n_{\mathrm{infl}}}\frac{d+2\delta}{\sqrt{2\delta d}}
    -\left(\frac12+\nu^{(-)}\right)\ln\frac{d+2\delta}{\sqrt{2\delta d\left(1+\nu^{(-)}\right)}}\nonumber\\
    -\frac{\left(\frac d2+\delta\right)^2}{2\delta d}
    -\frac12\big(1+\nu^{(-)}\big)+\ln\frac{d+2\delta}{\sqrt{2\delta
    d\nu^{(-)}}}.\quad\qquad\label{sum2}
\end{eqnarray}
Inserting (\ref{sum1}) and (\ref{sum2}) to Eqs.~(\ref{trefl-}) and
thereupon to (\ref{thetavolrefldef}), after a simple rearrangement
one is left with the final result
\begin{eqnarray}\label{appr-}
    \theta_\mathrm{v.r.}\approx-\theta_c\Bigg[1-\frac{2R_c}{R}\Bigg(\frac12\ln\frac{1}{{e \left(1-\nu^{(-)}\right)}}
    +\ln\frac{1}{{\nu^{(-)}}}\qquad\nonumber\\
    -\left(\nu^{(-)}+\frac12\right)\ln\frac{e}{{
    \left(1+\nu^{(-)}\right)}}+\big(1-\nu^{(-)}\big)\ln\frac{R}{R_c}\Bigg)\!\Bigg]\quad
\end{eqnarray}
(the definition for $\nu^{(-)}(b)$ is Eq.~(\ref{nu-def})). Note that
logarithmic asymptotics of this expression at $\nu^{(-)}\to0$ and at
$\nu^{(-)}\to1$ agrees with the general law (\ref{log-log}).

\begin{figure}
\includegraphics{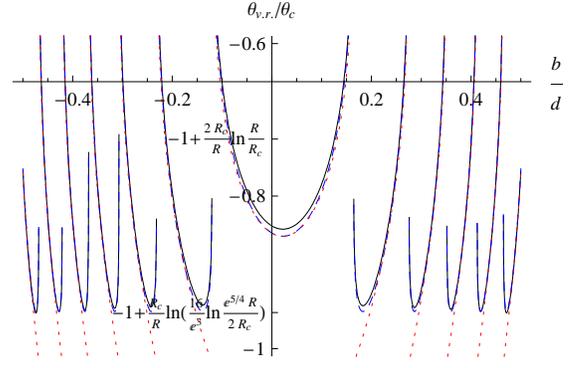}
\caption{\label{fig:approx-} Comparison of the exact formula
(\ref{thetavolrefldef}, \ref{trefl-}) for the negatively charged
particle indicatrix (black solid curve) with the approximation
(\ref{appr-}) (blue dashed curve) and approximation (\ref{z}) (red
dotted curve) for $R/R_c=25$. The central segment of the curves is
strongly $\theta_0$-dependent.}
\end{figure}

The exact (\ref{thetavolrefldef}) vs. approximate (\ref{appr-})
expressions for the indicatrix
$\theta_\mathrm{v.r.}\left(\nu^{(-)}(b)\right)$ are compared in
Fig.~\ref{fig:approx-}.

\subsubsection{Differential cross-section}\label{par:diff-cr-sect}

To deduce the observable differential cross-section from the
available indicatrix, we have again to issue from Eq.~(\ref{dlambda
dtheta}). To some extent, the same procedure as for positively
charged particles applies here, leading to representation
\begin{equation}\label{83}
  \frac{d\lambda}{d\theta_{\mathrm{v.r.}}} = \sum^2_{j=1}\left|\frac{d\nu^{(-)}}{d\theta_{\mathrm{v.r.}}}\right|\sum_m\frac1{|\partial\nu^{(-)}_j/\partial b|}\Bigg|_{
                                                                                                                                           b=b_m\left(\nu^{(-)}_j(\theta_{\mathrm{v.r.}})\right)
                                                                                                                                       },
\end{equation}
where $b_m\left(\nu^{(-)}\right)$ is the solution of
Eq.~(\ref{nu-def}), $\sum^2_{j=1}$ accounts for the existence of two
roots for equation
$\theta_{\mathrm{v.r.}}=\theta_{\mathrm{v.r.}}\left(\nu^{(-)}\right)$
with the function (\ref{appr-}). Again, upon averaging over
$\theta_0$ under condition (\ref{ineqs}) (cf. Appendix), one
obtains:
\begin{equation}\label{76}
    \frac
    1d\left\langle\frac{d\lambda}{d\theta_{\mathrm{v.r.}}}\right\rangle_{\theta_0}
    =\sum^2_{j=1}\left|\frac{d\nu^{(-)}_j}{d\theta_{\mathrm{v.r.}}}\right|,
\end{equation}
i. e., averaging over $b$ (and a tiny interval of $\theta_0$)
reduces to averaging over $\nu^{(-)}$ (transverse energy). Note that
when $\nu^{(-)}$ becomes a uniformly distributed random quantity in
a unit variation interval, for each given $b$, and thus
$\theta_{\mathrm{v.r.}}$, one can always unambiguously tell whether
$\nu^{(-)}=\nu^{(-)}_1$ or $\nu^{(-)}=\nu^{(-)}_2$, so the
probability normalization is conserved under conditions of summation
over the branches.

A technical distinction of the case with negative particles is that
equation (\ref{appr-}) can not be resolved with respect to
$\nu^{(-)}$ in an explicit and exact form. Of course, it can be
easily solved numerically; the differential cross-section so
evaluated is shown in Fig.~\ref{fig:diff-cross-}, by a solid curve.
On the other hand, it is also useful to pursue an analytic but
approximate approach, based on different approximations in different
regions of $\nu^{(-)}$, to which we yet pay some labor.

\subsubsection{Asymptotic evaluation of the final beam profile}

In Eq.~(\ref{appr-}) at typical $\nu^{(-)}$ the leading term is the
last one, where $\nu^{(-)}$ is multiplied by a large logarithm.
Besides that, in the domain of $\nu^{(-)}$ close to 1 the first
logarithm in (\ref{appr-}) becomes large, too, and a minimum of
function $\theta_{\mathrm{v.r.}}\left(\nu^{(-)}\right)$ develops,
corresponding to onset of a rainbow scattering. As for the second
logarithm in (\ref{appr-}), which raises at $\nu^{(-)}\to0$, it does
not lead to formation of a dependence
$\theta_{\mathrm{v.r.}}\left(\nu^{(-)}\right)$ minimum -- on the
contrary, it makes the dependence steeper, and in the area of its
significance the differential cross-section is small
(exponentially). So, for the differential cross-section description
it basically suffices to consider only two regions: the region where
the last term of (\ref{appr-}) dominates, and the region where the
last term and $\frac12\ln\frac1{e\left(1-\nu^{(-)}\right)}$ in
(\ref{appr-}) are competing. On the $\theta_{\mathrm{v.r.}}$ axis
the mentioned regions are adjacent, and conjointly they should give
almost the full picture of the differential cross-section variation.
For completeness, one may consider also a third, asymptotic region
of the the differential cross-section tail (orbiting region), where
the first and the second logarithms of (\ref{appr-}) dominate.

\paragraph{Rainbow region.}

The value of $\nu^{(-)}$ corresponding to the rainbow angle is to be
determined from condition
\begin{equation}\label{minimum}
    \frac{\partial\theta_{\mathrm{v.r.}}}{\partial\nu^{(-)}}\bigg|_{\nu^{(-)}=\nu^{(-)}_0}=0,
\end{equation}
which in application to expression (\ref{appr-}) gives
\begin{equation}\label{87}
    \frac{\nu_0^{(-)}}{1-\nu_0^{(-)2}}-\frac1{\nu^{(-)}_0}+\ln\left(1+\nu^{(-)}_0\right)=\ln\frac
    R{R_c}\gg1.
\end{equation}
The approximate solution of Eq.~(\ref{87}) is
\begin{equation}\label{}
    \nu^{(-)}_0\approx1-\frac1{2\ln\frac{e^{5/4}R}{2R_c}}+\mathcal{O}\left(\frac1{\ln^3\frac{e^{5/4}R}{2R_c}}\right).
\end{equation}

In vicinity of the found point $\nu^{(-)}_0$ we may expand function
$\theta_{\mathrm{v.r.}}\left(\nu^{(-)}\right)$ up to a quadratic
term
\begin{eqnarray}\label{y}
    \left(\frac{\theta_{\mathrm{v.r.}}}{\theta_c}+1\right)\!\frac R{2R_c}&\approx&\frac12\ln\left(\frac{16}{e^3}\ln\frac {e^{5/4}R}{2R_c}\right)\nonumber\\
    &+&\!\!\left(\nu^{(-)}\!-\nu^{(-)}_0\right)^2\!\ln^2\!\frac {e^{5/4}R}{2R_c}\nonumber\\
        &+&\!\!\mathcal{O}\left(\!\left(\nu^{(-)}-\nu^{(-)}_0\right)^3\ln^3\frac
        {e^{5/4}R}{2R_c}\right).\qquad
\end{eqnarray}
Now, expressing the pair of roots
$\nu^{(-)}_j\left(\theta_{\mathrm{v.r.}}\right)$ from the quadratic
equation (\ref{y}), one derives by formula (\ref{83}) the behavior
of the $\theta_0$-averaged differential cross-section in vicinity of
the rainbow angle:
\begin{equation}\label{1oversqtr}
    \frac{\theta_c}{d}\!\left\langle\!\frac{d\lambda}{d\theta_{\mathrm{v.r.}}}\!\right\rangle_{\theta_0}
    \approx\frac{\sqrt{R/2R_c}}{\ln\frac{e^{5/4}R}{2R_c}\sqrt{\frac{\theta_{\mathrm{v.r.}}}{\theta_c}+1-\frac{R_c}{R}\ln\left(\frac{16}{e^3}\ln\frac{e^{5/4}R}{2R_c}\right)}}.
\end{equation}
The domain of applicability of this approximation is determined from
Eq.~(\ref{y}) by demanding the third-order term to be small compared
with the second-order one:
\begin{equation}\label{cond-rainb}
    \left(\frac{\theta_{\mathrm{v.r.}}}{\theta_c}+\!1\!\right)\!\frac{R}{R_c}-\ln\!\left(\frac{16}{e^3}\ln\frac{e^{5/4}R}{2R_c}\right)\ll1\quad
    (\mathrm{rainbow\,region}).
\end{equation}

\begin{figure}
\includegraphics[width=87mm]{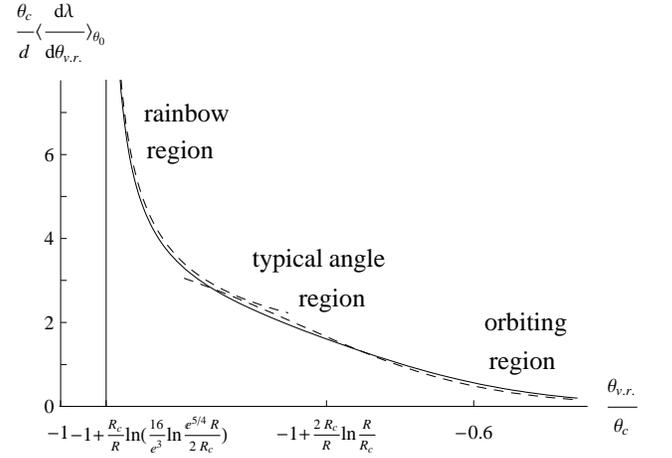}
\caption{\label{fig:diff-cross-} The $\theta_0$-averaged
differential cross-section of negatively charged particle
scattering, at $R/R_c=25$ (solid curve). The axes scales are chosen
so that areas under the curve is unity, as the total probability.
The left-hand dashed curve is evaluated by the explicit approximate
formula (\ref{1oversqtr}), the right-hand dashed curve -- by the
explicit approximate formula (\ref{1overLambert}).  It is observed
that those approximations actually overlap.}
\end{figure}

Function (\ref{1oversqtr}) is shown by the left-hand dashed curve in
Fig.~\ref{fig:diff-cross-}.

\paragraph{Typical angle region.}
On the other hand, if $\nu^{(-)}$ is not too close to 1, i. e. under
condition
\begin{equation}\label{cond-gen}
    \left(\frac{\theta_{\mathrm{v.r.}}}{\theta_c}+\!1\!\right)\!\frac{R}{R_c}-\ln\left(\frac{16}{e^3}\ln\frac{e^{5/4}R}{2R_c}\right)\gtrsim1\quad
    (\mathrm{typical\,angles})
\end{equation}
opposite to (\ref{cond-rainb}), all the terms in (\ref{appr-})
except the second logarithm (singular at $\nu^{(-)}\to0$) may
reasonably be approximated by their Taylor expansions up to linear
terms -- say, about the unit interval midpoint
$\nu^{(-)}_{\mathrm{mid}}=\frac12$ \footnote{More elegant
expressions, at the same time accommodating for a correct
asymptotics in the orbiting region, result when the Taylor expansion
is carried out about the point $\nu^{(-)}=0$, but unfortunately,
that approximation would be definitely less accurate at typical
$\theta_{\mathrm{v.r.}}$.}:
\begin{equation}\label{z}
    \theta_\mathrm{v.r.}\!\approx-\theta_c\!\left[1\!-\!\frac{2R_c}{R}\!\left(\ln\!\frac{\!\sqrt3R}{{\nu^{(-)}R_c}}\!
    -\!\frac{11}6\!-\!\nu^{(-)}\!\ln\!\frac{2R}{3e^{2/3}R_c}\!\right)\!\right]\!.
\end{equation}
Then, for determination of the inverse function
$\nu^{(-)}\left(\theta_{\mathrm{v.r.}}\right)$ one obtains the
following simplified equation:
\begin{equation}\label{eqq}
    \left(\frac{\theta_{\mathrm{v.r.}}}{\theta_c}+1\right)\frac
    R{2R_c}+\frac{11}6
    \approx\ln\!\frac{\!\sqrt3R}{\nu^{(-)}R_c}
    -\nu^{(-)}\ln\!\frac{2R}{3e^{2/3}R_c}.
\end{equation}
Here the r. h. s. is a monotonic function of $\nu^{(-)}$, so, in
contrast to the exact equation (\ref{appr-}), the simplified
equation (\ref{eqq}) has only one root (the second, lost root gives
a small contribution to the differential cross-section). The
solution to Eq.~(\ref{eqq}) expresses through the Lambert (product
log) function $W(s)$ defined as a solution to equation $\ln s=\ln
W+W$, and incorporated in many computational software packages:
\begin{equation}\label{jj}
    \nu^{(-)}\approx\frac1{\ln\frac{2R}{3e^{2/3}R_c}}
    W\!\left(\frac{\sqrt3 R \ln\frac{2R}{3e^{2/3}R_c}}{R_c}
    e^{-\frac{11}{6}-\frac{R}{2R_c}\left(\frac{\theta_{\mathrm{v.r.}}}{\theta_c}+1\right)}\right).
\end{equation}
(Mind that $\theta_{\mathrm{v.r.}}/\theta_c$ is negative and close
to $-1$). From the definition of $W$, its asymptotic behavior in
different regions derives as
\begin{equation}\label{}
   W(s)=\Bigg\{\begin{array}{c}
                              s-\mathcal{O}\left(s^2\right)\qquad s\ll1 \\
                            \ln\frac {s}{\ln \frac {s}{\ln \frac {s}{\ldots}}} \qquad\quad
                            s\gg1,
                          \end{array}
\end{equation}
and its derivative
\[
W'(s)=\frac1{s\left(1+1/W(s)\right)}.
\]
Therefore, over the typical angle region the $\theta_0$-averaged
differential cross-section is cast as
\begin{equation}\label{1overLambert}
    \frac{\theta_c}{d}\!\left\langle\!\frac{d\lambda}{d\theta_{\mathrm{v.r.}}}\!\right\rangle_{\theta_0}
    =\theta_c\left|\frac{d\nu^{(-)}}{d\theta_{\mathrm{v.r.}}}\right|
    \approx\frac R{2R_c\ln\frac
    {2R}{3e^{2/3}R_c}}\frac1{\left(1+1/W\right)},
\end{equation}
the argument of $W$ being the same as in Eq.~(\ref{jj}).

The full width of the differential cross-section as can be inferred
from Eq.~(\ref{1overLambert}), and is obvious already from
Eq.~(\ref{appr-}), amounts
\begin{equation}\label{width-}
    \triangle\theta_{\mathrm{v.r.}}\sim\theta_c\frac{2R_c}{R}\ln\frac{R}{R_c}\qquad(\mathrm{negatively\,charged\,particles}),
\end{equation}
which at $R/R_c>e^\pi\approx23$ is larger than width (\ref{width+})
for positively charged particles.

The behavior of function (\ref{1overLambert}) in region
(\ref{cond-gen}) is shown in Fig.~\ref{fig:diff-cross-} by the
right-hand dashed curve. It describes the actual distribution quite
accurately. In the orbiting region asymptotics, however,
Eq.~(\ref{1overLambert}) errs by a factor of
$e^{5/6}/\sqrt{3}\approx1.3$ (see Eq.~(\ref{orb}) below), so
applicability of (\ref{1overLambert}) is restricted by the bound
\begin{equation*}\label{}
    \frac{R}{2R_c}\left(\frac{\theta_{\mathrm{v.r.}}}{\theta_c}+1\right)\lesssim \ln\frac{R}{R_c}\qquad
    (\mathrm{typical\,angles})
\end{equation*}
in addition to (\ref{cond-gen}).

One may anticipate formation of a shoulder in the distribution
(\ref{1overLambert}) at sufficiently large $R/R_c$, since then the
argument of $W$ can be large and $1+1/W\to1$, making the
differential cross-section $\theta_{\mathrm{v.r.}}$-independent.
However, $W(s)$ reaches 5 ($\gg1$) only at $s\sim 700$, so a
shoulder in $\left\langle
d\lambda/d\theta_{\mathrm{v.r.}}\right\rangle_{\theta_0}$ may form
up only at $R/R_c\gtrsim100$.

\paragraph{Orbiting region.}
Finally, asymptotically large $\theta_{\mathrm{v.r.}}$ are generated
in the regions $\nu^{(-)}\to0$ and $\nu^{(-)}\to1$, where there are
logarithmically rising terms in the relation (\ref{appr-}). Examine
first the region $\nu^{(-)}\to0$. In this limit one may let
$\nu^{(-)}=0$ everywhere in (\ref{appr-}) except in
$\ln\frac1{\nu^{(-)}}$, and the reduced equation
\begin{equation*}\label{}
    \theta_\mathrm{v.r.}\simeq-\theta_c\left(1-\frac{2R_c}{R}\ln\frac{R}{eR_c\nu^{(-)}}\right)
\end{equation*}
is easily solved:
\begin{equation}\label{}
    \nu^{(-)}\simeq\frac{R}{eR_c}e^{-\frac{R}{2R_c}\left(\frac{\theta_{\mathrm{v.r.}}}{\theta_c}+1\right)}.
\end{equation}
Differentiating that expression with respect to
$\theta_{\mathrm{v.r.}}$, the differential cross-section asymptotics
results as
\begin{equation}\label{orb}
    \frac{\theta_c}{d}\!\left\langle\!\frac{d\lambda}{d\theta_{\mathrm{v.r.}}}\!\right\rangle_{\theta_0}
    \simeq\frac{R^2}{2eR^2_c}e^{-\frac{R}{2R_c}\left(\frac{\theta_{\mathrm{v.r.}}}{\theta_c}+1\right)}\quad
    (\mathrm{orbiting\,region}),
\end{equation}
holding if
\begin{equation*}\label{}
    \frac{R}{2R_c}\left(\frac{\theta_{\mathrm{v.r.}}}{\theta_c}+1\right)\gg \ln\frac{R}{R_c}\qquad
    (\mathrm{orbiting\,region}).
\end{equation*}

As for the contribution from region $\nu^{(-)}\to1$, which can be
treated in a completely analogous manner, it equals
\[
\frac{8R}{e^4R_c}e^{-\frac{R}{R_c}\left(\frac{\theta_{\mathrm{v.r.}}}{\theta_c}+1\right)},
\]
and is definitely negligible compared to (\ref{orb}), so
Eq.~(\ref{orb}) is the complete asymptotic result.

\subsubsection{Comparison with experiment}

The only measurement of negatively charged particle volume
reflection is \cite{Scandale-neg}. It gave, for Si (110)
orientation, $R/R_c\approx70$, the volume reflection angle mean
value
$\left|\left\langle\theta_{\mathrm{v.r.}}\right\rangle\right|=0.66\theta_c$.
This is significantly smaller than our expectation (and other
simulations, as quoted in \cite{rad})
$\left|\left\langle\theta_{\mathrm{v.r.}}\right\rangle\right|\approx\theta_c$
(yet minus $\sim8\%$ correction for finite $R_c/R$, which does not
matter anyway). Of course, if we treat $\left|F_{\max}\right|$ as an
adjustable parameter of the model, we might achieve agreement with
the experiment, but a physical recipe for this is yet lacking.

As for numerical calculation results for the final beam profile,
there is lack thereof for negatively charged particles at $R\gg R_c$
and free of multiple scattering (and with boundary conditions
congruent with typical experimental ones) \footnote{Paper
\cite{Maisheev} does not exhibit negative particle results for
$R>R_c$. In the pioneering work \cite{Tar-Vor} the boundary
conditions of are different from ours and from typical experimental
arrangements: the authors consider particles entering the crystal
face parallel to the bent atomic planes (a ``half" volume
reflection). That should produce a different final beam shape (with
no rainbow singularity and more particles in the ``orbiting" tail),
but nonetheless, Fig.~4b of \cite{Tar-Vor} is visually rather
similar to our Fig.~\ref{fig:diff-cross-}.}. We hope to see such
results in near future.

\section{High-energy passage limit (perturbative deflection)}\label{sec:pert-limit}

\begin{figure}
\includegraphics{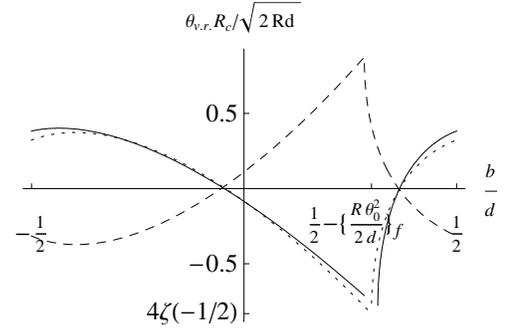}
\caption{\label{fig:high-en} Comparison of impact parameter
dependencies of the reflection angle for positively charged (solid
curve) and negatively charged (dashed curve) particles, at
$R/R_c=1/20$. Dotted curve -- approximation (\ref{Born}), for
positive particles. The functions displayed may be continuously
periodically extended beyond the interval
$\left(-\frac12,\frac12\right)$. The position of the fracture at
$\frac bd=\frac12-\left\{\frac{R\theta_0^2}{2d}\right\}_\mathrm{f}$
corresponds to tangency of the particle near-straight trajectory to
one of the bent atomic planes.}
\end{figure}

In conclusion, we will briefly comment on behavior of the function
$\theta_{\mathrm{v.r.}}(b)$ in the opposite, high-energy limit,
\begin{equation}\label{pert-limit}
\frac d{2\delta}=\frac{R}{R_c}\ll1.
\end{equation}
Thereunder, the deflection becomes perturbative (and better viewed
in Cartesian coordinates, without the reference to a centrifugal
force notion), and for positive and negative particles it must be
equal in magnitude but opposite in sign. That is confirmed by
Figs.~\ref{fig:theta(b)} and Fig.~\ref{fig:high-en}.

The specific expression for the dependence
$\theta_{\mathrm{v.r.}}(b)$ in this limit was obtained in
\cite{Bond-Shch} (Eqs.~(18-19)), in the Cartesian coordinate
framework:
\begin{subequations}\label{gh}
\begin{eqnarray}
    \theta_{\mathrm{v.r.}}\!\left(\theta_0,b\right)\!&\to&\!\theta_{\mathrm{Born}}\!\left(\theta_0,b\right)=\frac1E\int^\infty_{-\infty} dzF(b,z)\label{Born-integral}\\
    \!&=&\!\pm4\frac{\!\sqrt{2Rd}}{R_c}\zeta\!\left(\!-\frac12,\left\{\frac12+\frac{R}{2d}\theta_0^2+\frac bd\right\}_\mathrm{f}\right)\qquad\nonumber\\
    \label{Born}
\end{eqnarray}
\end{subequations}
(again, the braces indicate taking the fractional part). It involves
the Hurwitz (generalized Riemann) zeta-function at a negative value
of its first argument, which may be defined, e. g.,  as a contour
integral \cite{Eul-Macl}
\begin{equation}\label{Hurwitz-def}
    \zeta(\alpha,v)=\frac{\Gamma(1-\alpha)}{2\pi
    i}\int^{(0+)}_{-\infty}\frac{s^{\alpha-1}e^{vs}}{1-e^s}ds
\end{equation}
along a Hankel path \footnote{Conventionally, the Hankel path is
defined to begin in the complex $s$-plane at $-\infty$
($\mathrm{arg}\,s=-\pi $), encircle the origin in the positive
direction and return to $-\infty$ ($\mathrm{arg}\,s=+\pi $).}.
Function (\ref{Born}) is shown in Fig.~\ref{fig:high-en} by dotted
line. Note the identity
$\zeta\left(-\frac12,0\right)=\zeta\left(-\frac12,1\right)$
($\equiv\zeta(-\frac12)$), whereby function (\ref{Born}) is
everywhere continuous, but its derivative breaks at point $\frac
bd=\frac12-\left\{\frac{R\theta_0^2}{2d}\right\}_\mathrm{f}$.

The second argument of $\zeta$-function in (\ref{Born}) allows for a
physical interpretation \cite{Bond-Shch}:
\begin{equation}\label{brefl}
    b_{\mathrm{refl}}=b+\frac R2\theta_0^2
\end{equation}
is the particle impact parameter at a depth where the particle
straight trajectory becomes tangential to the bent atomic planes
(actually, $t_{\mathrm{refl}}$). Should one pass to
$b_{\mathrm{refl}}$ instead of $b$, the dependence on $\theta_0$
disappears completely:
\begin{equation}\label{}
    \theta_{\mathrm{v.r.}}\left(\theta_0,b\right)=\theta_{\mathrm{v.r.}}\left(b_{\mathrm{refl}}\right).
\end{equation}
This contrasts with the case $R\gg R_c$ investigated in the previous
section, where the $\theta_0$-dependence yet remained in a specific,
casually located rainbow peak.

Here we will not contemplate demonstrating from expressions
(\ref{trefl2-c},\ref{trefl-neg-2}) that the limit of
$\theta_{\mathrm{v.r.}}$ is indeed (\ref{Born}). We only note that
to this end one must implement significant cancellations between the
pair of terms under the sum sign. One also notes that the second
argument of $\zeta$-function in (\ref{Born}) is just the limit of
$\nu^{(+)}$:
\begin{equation}\label{}
    \underset{\delta/d\to\infty}\lim\nu^{(+)}=\left\{\frac12+\frac{R}{2d}\theta_0^2+\frac
    bd\right\}_\mathrm{f}.
\end{equation}
For negative particles the route to (\ref{Born}) is a bit more
intricate. In Fig.~\ref{fig:high-en} we numerically compare
approximation (\ref{Born}) with the exact result -- the agreement is
quite convincing.

The comparison with the perturbative scattering pattern may also
give an insight into the origin of the volume reflection phenomenon.
The average of function (\ref{gh}) over the impact parameter $b$
turns out to be strictly zero (see \cite{Bond-Shch}) -- hence, in
the high-energy limit signatures of the volume reflection completely
disappear. That owes to the fact that $\int^{d/2}_{-d/2} db F(b,
z)\equiv0$ at any given $z$ (regardless of the crystal bend). In
contrast, at $R\gtrsim R_c$ the trajectory may \emph{not} be viewed
as merely straight \footnote{In the literature, sometimes, one meets
an interpretation of the volume reflection phenomenon right in terms
of a straight trajectory tangential to the bent crystalline planes
in some point. Such an interpretation, although satisfactory for a
symmetric installation of crystals with a large bending angle
($\sim2\theta_0\gg\theta_c$), may be misleading for understanding of
the underlying particle dynamics.}, and the particle distribution is
non-uniform over the crystal volume. In fact, ``shadowed regions"
not filled by the particles may appear at the inner side of bent
potential ridges, in which the force acts in the positive direction.
The deficit of a positively directed force on the beam then leads to
the negative sign of the particle beam mean deflection angle.

\section{Experimental objectives and parameter
optimization}\label{sec:optimiz} The alleged application of volume
reflection is for high-energy particle beam extraction from
accelerator beamlines, when that is not manageable locally with
laboratory magnets (intra-crystal forces are much stronger).
Thereat, the beam parameters (energy and angular divergence) are
definite, while the crystal parameters have to be optimized in order
to attain a suitable deflection quality.

\subparagraph{Control of beam divergence accompanying the
deflection.}

The mean deflection angle
\begin{eqnarray}\label{98}
\left|\theta_{\mathrm{v.r.}}\right|\sim\theta_c,\quad
\theta_c=\sqrt{\frac{2V_0}{E}}\approx21\, \mu\mathrm{
rad}\sqrt{\frac{100\,\mathrm{GeV}}{E}}\quad\\
(\mathrm{for\,Si\,(110)})\qquad\qquad\qquad\qquad\nonumber
\end{eqnarray}
for a silicon crystal of orientation (110) depends only on the
particle energy. The next parameter to care about is the angular
spread acquired by the beam at the exit from the crystal. Neglecting
the incoherent multiple scattering (to be estimated below), the
spread $\triangle\theta_{\mathrm{v.r.}}$, given by equations
(\ref{width+}) and (\ref{width-}), for a given beam energy and
crystal material depends only on the crystal bending radius. To
derive a criterion for beam complete deflection, one may demand that
the bulk of the dispersed beam (its both ``edges") be deflected to
the same side. That implies:
\begin{equation}\label{R>4R_c}
R>4R_c\qquad (\mathrm{for\, positively\, charged\, particles})
\end{equation}
and
\begin{equation}\label{R>4R_clog}
R>2R_c\ln\frac R{R_c}. \qquad (\mathrm{for\,negatively\,charged\,
particles})
\end{equation}
Relation (\ref{R>4R_c}) with the factor 4 was found previously by
Maisheev \cite{Maisheev} based on numerical simulation studies for
protons. Our paper, thereby, offers a formal justification for that
empirical relation, although within a framework of a simplified
model. The emerging ratio
\begin{equation}\label{qdef}
        \mathsf{q}=\frac{\max\left|\theta_{\mathrm{v.r.}}\right|}{\triangle\theta_{\mathrm{v.r.}}}=\Bigg\{\begin{array}{c}
                                                 \frac{R}{4R_c} \qquad\quad \mathrm{for\, pos.\, charged\, particles}\\
                                                 \frac{R}{2R_c\ln\frac{R}{R_c}} \quad \mathrm{for\, neg.\, charged\, particles}
                                               \end{array}
\end{equation}
quantifies the steering quality. But for the deflection to be neat,
one should have a substantial $\mathsf{q}$
\begin{equation}\label{}
    \mathsf{q}\gtrsim3\qquad(\mathrm{neat\, deflection}),
\end{equation}
entailing for the crystal curvature radius
\begin{equation}
R\gtrsim10\div15R_c.
\end{equation}
In demonstrational experiments \cite{Scandale} the latter
requirement was marginally satisfied.

\subparagraph{Sufficient crystal thickness.}

The crystal thickness ($L$), it to be kept as low as possible in
order to minimize the multiple scattering. However, a lower bound
for $L$ results from the requirement that the volume reflection has
space to develop, i.~e., the ``thick crystal limit" holds, in the
sense of Sec.~\ref{subsec:thick-cryst}. Thereto, the crystal bending
half-angle $\frac{L}{2R}$ needs be larger than the critical angle
$\theta_c$, which is $L$-independent. Thus,
\begin{eqnarray}\label{Lmin}
L\gg
2R\theta_c=\frac{R}{R_c}d\sqrt{\frac{E}{2V_0}}\approx\frac{R}{4R_c}3.6\,\mu
\mathrm{m}\sqrt{\frac{E}{\mathrm{GeV}}}\quad\\
(\mathrm{thick\,crystal,\,volume\,reflection\,saturation}).\quad\nonumber
\end{eqnarray}
If we regard here $\frac{R}{4R_c}=\mathsf{q}$ as fixed, in general
$L$ grows with the energy. At $E\sim10^2\div10^3\,\mathrm{GeV}$
(RHIC, Tevatron) the minimal thickness amounts only to
$(0.036\div0.1\,\mathrm{mm})\mathsf{q}$. Note that crystals as thin
as $30\,\mathrm{mm}$ are manufacturable (as described in
\cite{Guidi}). For $E\sim10\,\mathrm{TeV}$ (LHC) and
$\mathsf{q}\sim3$ minimal $L$ reaches the value of $1\,\mathrm{mm}$.

\subparagraph{Reduction of multiple scattering.}

To quantify the impact of incoherent, random multiple scattering, we
have to evaluate the characteristic ratio \footnote{For the (rms,
plane) multiple scattering angle upon the particle over-barrier
passage we crudely apply a formula for the scattering angle in an
\emph{amorphous} target made of the same material (silicon):
$\theta^{\mathrm{mult}}_x=\frac{13.6\,\mathrm{MeV}}{\sqrt2
E}\sqrt{\frac{L}{93.6\,\mathrm{mm}}}$ (as quoted in \cite{Amsler}).}
\begin{eqnarray}\label{mult-c}
    \frac{\theta^{\mathrm{mult}}_x}{\theta_c}&\approx&\frac{13.6\,\mathrm{MeV}}{\sqrt2
    E}\sqrt{\frac{L}{93.6\,\mathrm{mm}}}\sqrt{\frac{E}{2V_0}}\nonumber\\
    &\approx&0.47\sqrt{\frac{100\,\mathrm{GeV}}{E}\frac{L}{\mathrm{mm}}}\qquad
(\mathrm{Si\,(110)}).\quad
\end{eqnarray}
If the latter square root does not exceed unity (in experiments
\cite{Scandale} it is $\approx1$), the multiple scattering does not
spoil coherent beam deflection. We conclude that for
$E\gtrsim100\,\mathrm{GeV}$ crystal lengths up to $2\,\mathrm{mm}$
are multiple scattering safe, i.~e.,
$\theta^{\mathrm{mult}}_x/\theta_c<1$.


Besides aggregate deflection, it would be interesting to
experimentally investigate the intrinsic volume-reflected beam
\emph{shape}, and in particular to check the shape dependence on the
particle charge sign (cf. Figs.~\ref{fig:diff-cross+} and
\ref{fig:diff-cross-}). The main problem here is that at
$\mathsf{q}\gg1$ the final beam half divergence
$\frac12\triangle\theta_{\mathrm{v.r.}}$ is $2\mathsf{q}$ times
smaller than the mean deflection angle, and so is sooner overtaken
by the multiple scattering, making the profile Gaussian and particle
charge sign independent. For this not to happen, one needs condition
\begin{equation}\label{req}
    \frac{\theta^{\mathrm{mult}}_x}{\frac12\triangle\theta_{\mathrm{v.r.}}}\sim
    2\mathsf{q}\frac{\theta^{\mathrm{mult}}_x}{\theta_c}\ll1,
\end{equation}
with $\theta^{\mathrm{mult}}_x/\triangle\theta_{\mathrm{v.r.}}$ to
be inferred from (\ref{mult-c}). Hence, for the present purpose we
should not strive for large $\mathsf{q}$, granted that the final
beam profile is not very sensitive to $\mathsf{q}$ at
$\mathsf{q}>1$. So, $\mathsf{q}\simeq1.2$ seem to be good enough.
Equally well, in order to raise the angular resolution we should use
moderate energies. Say, $E=50\,\mathrm{GeV}$ is ultra-relativistic
enough. To reduce multiple scattering, we can take a thin crystal
with $L=30\,\mu\mathrm{m}$, which marginally satisfies (\ref{Lmin}).
This gives $\theta_c\approx3\cdot10^{-5}\,\mathrm{rad}$,
$\triangle\theta_{\mathrm{v.r.}}\approx2.5\cdot10^{-5}\,\mathrm{rad}$,
and
$\frac{\theta^{\mathrm{mult}}_x}{\frac12\triangle\theta_{\mathrm{v.r.}}}\approx
0.3$ (small enough). But one has to control initial particle impact
angles with an accuracy a few times better than
$\triangle\theta_{v.r.}$. This may be difficult to achieve via
initial beam collimation alone, so one may need to apply event
selection procedures (cf. \cite{Scandale-neg}).


\section{Summary}\label{sec:summary}

Based on the model of a purely parabolic continuous potential in a
bent crystal, we have gained a lot of information about the volume
reflection phenomenon, for cases of positively and negatively
charged particles. First, we have obtained an explicit expression
(\ref{rn}) for particle trajectories. From the solution for
trajectory, in particular, we have derived the particle final
deflection angle as a function of the particle impact parameter and
energy, in form of sums (\ref{trefl2-c}, \ref{trefl-neg-2}).
Asymptotic behavior of those sums at $R\gg R_c$ was explored, and
asymptotic values for the volume reflection angle were found. They
equal: $-\frac\pi2\theta_c$ for positive particles, and $-\theta_c$
for negative particles. This agrees within $\sim20\%$ with the
existing results of numerical simulation using more realistic
continuous potentials \cite{Tar-Vor,Maisheev} and with experiment
for positive particles \cite{Scandale} (though there is an
indication of worse agreement for negative particles
\cite{Scandale-neg}). 20\% is about the same accuracy as for
approximating the continuous potential by a parabola. Yet we have
evaluated the next-to-leading order correction in parameter $R_c/R$,
which depends on the impact parameter, and, by averaging over impact
parameters, we determined asymptotic shape of the final beam. This
in particular yields the mean volume reflection angle dependence on
$R_c/R$, which appears to be linear -- in general agreement with
experiment \cite{Scand-linear} (see Fig.~\ref{fig:fit}).

In course of investigation of the final beam shape, we have
discovered various singularities in its profile, which moreover are
particle charge dependent. First of all, we had to deal with the
problem that, in principle, the final beam profile may contain a
visible admixture of boundary dependent effects (``randomly" located
peaks). However, we have proved the statement that boundary effects
get completely erased in the differential cross-section averaged
over a tiny interval of incident angles $\theta_0$ (condition
(\ref{ineqs})), or, analogously, due to a bit of multiple scattering
before the volume reflection region. Therewith, the averaging over
impact parameters becomes equivalent to averaging over parameters
$\nu^{(\pm)}$ (i.~e., transverse energy), and we were able to
analytically deduce the final beam profile for positive and for
negative particles. For negatively charged particles it is
asymmetric, exhibiting a spike on its outer edge, corresponding to
the rainbow scattering, and an exponential tail on the inner side,
corresponding to orbiting (Fig.~\ref{fig:diff-cross-}). For positive
particles, the final beam has a rectangular profile
(Fig.~\ref{fig:diff-cross+}). But in actual practice, with the
account of continuous potential smearing in vicinity of the atomic
planes, one expects appearance of a weak rainbow spike and orbiting
tail for positive particles, as well.

Towards practical applications and further experimental
investigations, we have made a few numerical estimates. They
indicate that for usage of a bent crystal as a coherent beam
deflector, one needs a relation between the main parameters
\[
\frac{L}{1\,\mathrm{mm}}<\frac E{100\,\mathrm{GeV}}<\frac
R{\mathrm{m}},\,\left(\frac{20\,\mu\mathrm{rad}}{\sigma_0}\right)^2
\]
($\sigma_0$ is the r.m.s. angular deviation in the initial beam).
The better those inequalities are met, the higher is the deflection
quality. If one becomes interested in investigation of the final
beam intrinsic shape, generated by the continuous potential alone,
those inequality must be satisfied strongly, but minding existence
of technical lower limits for $L$ and $\sigma_0$. This suggests an
optimal energy about 50 GeV; experiments are to be carried out
simultaneously with particles of both charge signs ($e^\pm$,
$\pi^\pm$).



There are many respects in which the model solution described herein
can be improved. First of all, it is straightforward to add to the
simple parabolic potential a second parabolic section -- either to
round off the potential in vicinity of atomic planes, or to describe
the potential of Si crystal in (111) planar orientation, which is of
practical importance, too. As a next step -- at least a perturbative
account of incoherent scattering processes is desirable. But at the
same time, even in the present form, the theory (trajectories
derived in Sec.~\ref{sec:1stinterv}) seems suitable, e. g., for
study of electromagnetic radiation emitted by a volume-reflected
particle.


\appendix
\section{Formal procedure of $\theta_0$-averaging}
In Sec.~\ref{sec:pos-part} we had obtained, for positively charged
particles, the scattering differential cross-section in form
\begin{equation}\label{A1}
    \frac{d\lambda}{d\theta_{\mathrm{v.r.}}}=\frac{R}{2\pi\theta_c
    R_c}\sum_m\frac1{\left|d\nu^{(+)}/db\right|_{b=b_m\left(\nu^{(+)}(\theta_{\mathrm{v.r.}}),\theta_0\right)}}.\quad
\end{equation}
Our objective now is to straightforwardly compute the sum involved
hereat for the specific function $\nu^{(+)}(b,\theta_0)$ given by
Eq.~(\ref{nu+def}), first for an arbitrary $\theta_0$, and then
average it over $\theta_0$, in order to justify our assertion that
combined averaging over $b$ and $\theta_0$ (within tiny interval
(\ref{ineqs})) is equivalent to averaging over $\nu^{(+)}$.

To begin with, let us find the roots $b_m$ explicitly.
Eq.~(\ref{nu+def}) is equivalent to
\begin{equation}\label{nu+m}
    \frac{\tau^2\theta_0^2+(b+\delta)^2-\left(\frac d2-\delta\right)^2}{2\delta
    d}=\nu^{(+)}+m
\end{equation}
with $m$ an integer, solution of which is straightforward:
\begin{equation}\label{bm}
    b_m(\nu^{(+)},\theta_0)=-\delta\pm\sqrt{2\delta d\left(\nu^{(+)}+m\right)+\!\left(\frac
    d2-\delta\right)^2\!-\tau^2\theta_0^2}.
\end{equation}
Here the sequence of $m$ begins with a smallest integer $m_0$ at
which the radicand in (\ref{bm}) is yet positive, viz.
\begin{equation}\label{m0}
    m_0=\left\lfloor\frac{\tau^2\theta_0^2-\left(\frac d2-\delta\right)^2}{2\delta
    d}-\nu^{(+)}\right\rfloor+1.
\end{equation}
The upper limit of $m$ in Eq.~(\ref{bm}) equals to the largest
integer at which yet $b<\frac d2$ (for a branch with the ``+" sign
in front of the root in Eq.~(\ref{bm})), and $b>-\frac d2$ (for a
branch with the ``$-$" sign). That yields, correspondingly, values
\begin{eqnarray}
  m_{\max1} &=& \left\lfloor\frac{\tau^2\theta_0^2}{2\delta
    d}-\nu^{(+)}\right\rfloor, \label{mmax1}\\
  m_{\max2} &=& m_{\max1}+1.\label{mmax2}
\end{eqnarray}
The number of terms in the sums from (\ref{m0}) to (\ref{mmax1},
\ref{mmax2}) is large:
\begin{equation*}
    m_{\max1,2}-m_0=\frac{d}{8\delta}+\mathcal{O}(1)\equiv\frac{R}{4R_c}+\mathcal{O}(1)\gg1.
\end{equation*}

Next, values of derivative $\frac{d\nu^{(+)}}{db}$ in points $b_m$
are easily evaluated, noticing that the fractional part operator in
$\nu^{(+)}$ is inconsequential for derivatives \footnote{Strictly
speaking, differentiation of finite discontinuities will give
$\delta$-functional terms, but they will be imperceptible when
inserted to the \emph{denominator} of equation (\ref{dlambda
dtheta}).}. Differentiating (\ref{nu+m}) gives
\begin{eqnarray}\label{}
    \frac{d\nu^{(+)}}{db}\bigg|_{b=b_m}=\frac{b_m+\delta}{\delta d}.
\end{eqnarray}
Identity
\begin{equation}\label{}
    \frac1{\left|d\nu^{(+)}/db\right|_{b=b_m}}=\left|\frac{\partial b}{\partial m}\right|
\end{equation}
suggests that upon substitution to (\ref{A1}) one may expect
\begin{eqnarray}\label{A9}
    \left(\sum_{m=m_0}^{m_{\max1}}+\sum_{m=m_0}^{m_{\max2}}\right)\frac1{\left|d\nu^{(+)}/db\right|_{b=b_m}}\approx2\sum_{m=m_0}^{m_{\max1}}\frac{\partial b}{\partial m}\qquad \nonumber\\
    \approx 2\int_{0}^{d/2}db=d,\quad
\end{eqnarray}
as we anticipated in Sec.~\ref{sec:pos-part}, but to accommodate the
dependence on $\theta_0$, we need to carry out the calculation more
precisely.

Through (\ref{bm}, \ref{A1}), the differential cross-section assumes
the form
\begin{eqnarray}\label{diff-cross-cusp}
    \frac{d\lambda(\theta_{\mathrm{v.r.}},\theta_0)}{d\theta_{\mathrm{v.r.}}\!}\!=\frac{R}{2\pi\theta_cR_c}\quad\qquad\qquad\qquad\qquad\qquad\qquad\nonumber\\
    \cdot\left(\sum_{m=0}^{m_{\max1}}\!+\!\sum_{m=0}^{m_{\max2}}\right)\!\frac{\delta d}{\!\sqrt{2\delta d\left(\nu^{(+)}\!+m\right)\!+\left(\frac
    d2\!-\delta\right)^2\!-\!\tau^2\theta_0^2}}\nonumber\\
    \approx\frac{d}{2\pi\theta_c}\sqrt{\frac{R}{R_c}}\sum_{m=0}^{R/4R_c}\frac1{\sqrt{m+\alpha\left(\nu^{(+)}(\theta_{\mathrm{v.r.}}),\theta_0\right)}}\qquad\qquad
\end{eqnarray}
with
\begin{eqnarray}
  \alpha\!\left(\nu^{(+)}(\theta_{\mathrm{v.r.}}),\theta_0\right)\! &=&\! 1- \left\{\!\frac{\tau^2\theta_0^2-\left(\!\frac d2-\delta\right)^2}{2\delta
    d}-\nu^{(+)}\!\right\}_{\mathrm{f}}\nonumber\\
   &\equiv&\!\left\{\frac{\left(b\left(\nu^{(+)}(\theta_{\mathrm{v.r.}}),\theta_0\right)+\delta\right)^2}{2\delta
   d}\right\}_{\mathrm{f}}\!.\qquad\,\,\,
\end{eqnarray}
Outside the interval
\begin{equation}\label{theta-interval}
    -\frac\pi2\theta_c<\theta_{\mathrm{v.r.}}<-\frac\pi2\theta_c\left(1-\frac{4R_c}{R}\right),
\end{equation}
there are no roots to equation $\nu^{(+)}(\theta_{\mathrm{v.r.}})$,
so the differential cross-section vanishes as an empty sum.

At a large upper limit the sum in (\ref{diff-cross-cusp}) grows as
$\sqrt{R/R_c}$, whereas the difference
\begin{equation}\label{zeta12}
    \sum_{m=0}^{R/4R_c}\frac1{\sqrt{m+\alpha}}-\sqrt{\frac
    R{R_c}}\underset{R/R_c\to\infty}\longrightarrow\zeta\left(\frac12,\alpha\right)
\end{equation}
tends to a finite limit (it may be categorized as the Hurwitz, or
generalized Riemann, zeta-function, see also definition
(\ref{Hurwitz-def})). Thereby, we may cast (\ref{diff-cross-cusp})
as
\begin{equation}\label{A14}
    \frac{d\lambda(\theta_{\mathrm{v.r.}},\theta_0)}{d\theta_{\mathrm{v.r.}}\!}\approx\frac{d}{2\pi\theta_c}{\frac{R}{R_c}}\left(1+\sqrt{\frac{R_c}{R}}\zeta\left[{\frac12},\alpha(\theta_{\mathrm{v.r.}},\theta_0)\right]\!\right).
\end{equation}
Here $\sqrt{R_c/R}\ll1$, and the unity in parentheses (\ref{A14})
corresponds to the result anticipated in (\ref{A9}). But one should
take care that at $\alpha\to0$ function
$\zeta\left(\frac12,\alpha\right)$ blows up as
$\frac1{\sqrt{\alpha}}$. So, the correction in the parentheses in
(\ref{A14}) can not be regarded as everywhere small.

Now, we turn to the issue that $\alpha$ is $\theta_0$-dependent.
When $\theta_0$ varies (at $\theta_{\mathrm{v.r.}}$ fixed) even in a
narrow interval (see (\ref{11})), $\alpha$ uniformly and repeatedly
scans its definition interval from 0 to 1. Hence, the averaging over
$\theta_0$ is equivalent to the integration over $\alpha$ from 0 to
1. By virtue of the property
\begin{equation}\label{int-alpha-zero}
    \int_0^1d\alpha\zeta\left(\frac12,\alpha\right)\equiv0,
\end{equation}
checkable from definition (\ref{zeta12}), we have
\begin{eqnarray}\label{}
    \left\langle\frac{d\lambda}{d\theta_{\mathrm{v.r.}}}\right\rangle_{\theta_0}=\int_0^1d\alpha\frac{d\lambda}{d\theta_{\mathrm{v.r.}}}\quad\qquad\qquad\qquad\qquad\qquad\qquad\nonumber\\
    =\frac{Rd}{2\pi\theta_c
    R_c}\Theta\left(\theta_{\mathrm{v.r.}}\!+\frac\pi2\theta_c\right)\Theta\left(\!-\theta_{\mathrm{v.r.}}-\frac\pi2\theta_c\left(1-\frac{4R_c}{R}\right)\!\right)\!.\nonumber
\end{eqnarray}
Thereby, we arrive at equation (\ref{diff-cross+aver}), without any
corrections $\sim\sqrt{R_c/R}$, q.e.d.

For negatively charged particles there arises a sum similar to that
of (\ref{A1}), though the dependence $\nu^{(-)}(b)$ differs a bit
from $\nu^{(+)}(b)$. Nonetheless, the averaging procedure is
completely analogous, the non-averaged differential cross-section
equals (\ref{76}) times the parentheses factor of (\ref{A14}). Upon
the $\theta_0$-averaging, through (\ref{A14}) again, we arrive at
Eq.~(\ref{76}).

\end{document}